\documentclass{jpp}
\usepackage{bm}
\usepackage{graphicx}
\usepackage{epstopdf, epsfig}
\usepackage{hyperref}
\usepackage{color}
\usepackage{amsmath,amssymb,latexsym}
\usepackage{natbib}
\usepackage{enumerate}   

\newcommand{\beq}[0]{\begin{equation}}
\newcommand{\eeq}[0]{\end{equation}}

\shorttitle{Particle Acceleration in Magnetic Reconnection}
\shortauthor{D. A. Uzdensky}

\title{Relativistic Nonthermal Particle Acceleration in Two-Dimensional Collisionless Magnetic Reconnection}

\author{Dmitri A. Uzdensky\aff{1}
  \corresp{\email{uzdensky@colorado.edu}}}

\affiliation{\aff{1}Center for Integrated Plasma Studies, Physics Department, 
   390 UCB, University of Colorado, Boulder, CO 80309, USA}

\begin{document}

\maketitle

\date{\today}


\begin{abstract}
Magnetic reconnection, especially in the relativistic regime, provides an efficient mechanism for accelerating relativistic particles and thus offers an attractive physical explanation for nonthermal high-energy emission from various astrophysical sources. I present a simple analytical model that elucidates key physical processes responsible for reconnection-driven relativistic nonthermal particle acceleration (NTPA) in the large-system, plasmoid-dominated regime in two dimensions. The model aims to explain the numerically-observed dependencies of the power-law index $p$ and high-energy cutoff $\gamma_c$ of the resulting nonthermal particle energy spectrum $f(\gamma)$ on the ambient plasma magnetization~$\sigma$, and (for $\gamma_c$) on the system size~$L$.  In this self-similar model, energetic particles are continuously accelerated by the out-of-plane reconnection electric field $E_{\rm rec}$ until they become magnetized by the reconnected magnetic field and eventually trapped in plasmoids large enough to confine them. The model also includes diffusive Fermi acceleration by particle bouncing off rapidly moving plasmoids. I argue that the balance between electric acceleration and magnetization controls the power-law index, while trapping in plasmoids governs the cutoff, thus tying the particle energy spectrum to the plasmoid distribution.  
\end{abstract}





\section{Introduction}
\label{sec-intro}

Nonthermal acceleration of relativistic particles is one of the most outstanding and important problems in theoretical plasma astrophysics. 
Vast numbers of astrophysical sources, scattered throughout our Galaxy and beyond, shine to us with powerful outbursts of high-energy (X-ray and gamma-ray) radiation.  
This radiation, routinely observed to reach into MeV, GeV, and, in some systems, TeV energy ranges, indicates that the emitting charged particles (electrons and, in some systems, positrons) are ultra-relativistic, with Lorentz factors $\gamma \gg 1$. 
It is therefore not surprising that most classes of the observed gamma-ray astrophysical sources are associated with relativistic objects --- neutron stars (NSs) and black holes (BHs) --- and their relativistic outflows.  The most notable examples are pulsar magnetospheres and pulsar wind nebulae (PWN); magnetars; accretion-disk coronae (ADCe) and radiatively-inefficient accretion flows (RIAFs) of both stellar-mass BHs in Galactic X-Ray Binaries (XRBs) and supermassive BHs (SMBHs), e.g., in active galactic nuclei (AGN), including the {\it EHT} sources M87 and Sgr~A*; BH-powered relativistic jets emanating from XRBs in their low-hard spectral state and from AGN, including blazars, as well as the jet-fed giant radio-lobes; and gamma-ray bursts (GRBs), both long GRBs produced by core-collapse of massive stars and in short GRBs produced by in BH-NS or NS-NS mergers (including gravitational waves events like~GW170817).%
\footnote {Not all GeV and TeV sources, however, are directly associated with, or powered by, NSs or~BHs; in particular, an important class of gamma-ray sources are nonrelativistic shocks driven through the interstellar medium by powerful blast waves in supernova (SN) remnants (SNRs).}

The high-energy (gamma-ray) radiation from many of these sources, both quasi-stationary (persistent) and violently flaring (bursty), is often observed to have nonthermal spectra, characterized by extended power laws covering several orders of magnitude in photon energy. Because the most common emission mechanisms capable of emitting photons in this energy range --- synchrotron, inverse-Compton (IC), and curvature --- involve just single individual particles interacting with ambient background magnetic or soft-photon fields and are characterized by simple power-law relationships between the emitting particle's energy $\gamma m_e c^2$ and the resulting photon energy $\epsilon_{\rm ph}$ (e.g., $\epsilon_{\rm ph} \sim \gamma^2$ for synchrotron and IC), the power-law radiation spectrum implies a power-law energy distribution, $f(\gamma) \sim \gamma^{-p}$, of the emitting particles.%
\footnote{In addition to inferring astrophysical nonthermal particle acceleration through the nonthermal electromagnetic radiation that these particles produce, we have a direct evidence of nonthermal relativistic particles pervading our Galaxy---cosmic rays.}

In fact, nonthermal spectra are so prevalent in gamma-ray (and some hard X-ray) sources that 
one is lead to think that they are more of a norm rather than an exception. This, by itself, is not surprising; the very fact that we can observe nonthermal radiation of such high energy, i.e., that the gamma-ray photons can escape from the system and its immediate surroundings, often suggests that we are dealing with environments with a modest or small optical depth to Compton scattering. This, in turn, means that the plasma density is low and, since the Coulomb cross-section for relativistic particles becomes comparable to the Thomson cross-section, that the plasma is collisionless. Therefore, there is no a priori reason for the energy conversion processes that energize the plasma and power the emission to produce thermal particle populations.  Nevertheless, it is an important and interesting intellectual challenge to understand the concrete physical mechanisms at work driving nonthermal particle acceleration (NTPA); in particular, one would like to build a predictive theory, capable of explaining the key characteristics of NTPA, such as its overall energy efficiency, the power-law index~$p$, and the high-energy cutoff~$\gamma_c$. 

Charged particle acceleration requires work to be done on the particle by an electric field, and, for the resulting energy gain to be large, one has to have strong electric fields coherent over substantial distances.  Since highly conducting, collisionless astrophysical plasmas tend to efficiently screen electric fields that are parallel to the magnetic field in the plasma comoving frame, most of the required macroscopic electric fields would have to be ideal magnetohydrodynamic (ideal-MHD), motional (${\bf u} \times {\bf B}$) electric fields associated with rapid bulk motions of magnetized plasmas.  For this reason, most of the astrophysically relevant NTPA mechanisms are based on some dynamic, often rather violent, plasma processes. Namely, the three such candidate processes that are most often invoked in theoretical models of relativistic NTPA are collisionless shocks, turbulence, and magnetic reconnection.  It is now generally believed that all three provide plausible, viable mechanisms for particle acceleration under different conditions.  They all have been extensively studied as NTPA drivers both analytically \citep{Bulanov_Sasorov-1976, Blandford_Ostriker-1978, Blandford_Eichler-1987, Schlickeiser-1989, Chandran-2000, Larrabee_etal-2003, Giannios-2010} and numerically. In particular, their viability as relativistic particle accelerators  has been recently established in first-principles particle-in-cell (PIC) kinetic plasma simulations 
[see, e.g., \cite{Spitkovsky-2008, Sironi_Spitkovsky-2011a, Sironi_Spitkovsky-2011b, Caprioli_Spitkovsky-2014} for shocks; 
\cite{Zhdankin_etal-2017, Zhdankin_etal-2018b, Zhdankin_etal-2019, Comisso_Sironi-2018, Comisso_Sironi-2019, Wong_etal-2020} for turbulence; and \cite{Zenitani_Hoshino-2001, Zenitani_Hoshino-2005, Zenitani_Hoshino-2007, Zenitani_Hoshino-2008, Jaroschek_etal-2004a, Lyubarsky_Liverts-2008, Liu_etal-2011, Bessho_Bhattacharjee-2012, Cerutti_etal-2013, Cerutti_etal-2014a, Cerutti_etal-2014b, Sironi_Spitkovsky-2014, Melzani_etal-2014b, Cerutti_etal-2015, Sironi_etal-2015, Sironi_etal-2016, Guo_etal-2014, Guo_etal-2015, Guo_etal-2016, Guo_etal-2019, Nalewajko_etal-2015, Werner_etal-2016, Werner_Uzdensky-2017, Werner_etal-2018, Ball_etal-2018, Werner_etal-2019, Schoeffler_etal-2019, Hakobyan_etal-2019, Mehlhaff_etal-2020, Hakobyan_etal-2020} for relativistic magnetic reconnection (see also \citet{Hoshino_Lyubarsky-2012, Kagan_etal-2015} for recent reviews)]. There has also been a notable body of influential theoretical and numerical work on nonrelativistic NTPA in reconnection, e.g., in the context of solar and space physics \citep{Drake_etal-2006,  Drake_etal-2010, Drake_etal-2013, Oka_etal-2010,  Dahlin_etal-2014, Dahlin_etal-2015, Dahlin_etal-2016, Dahlin_etal-2017, Li_etal-2015, Li_etal-2017}. 


The main goal of the present paper is to develop an analytical theory of relativistic nonthermal particle acceleration (NTPA) in one of these processes --- collisionless magnetic reconnection. In particular, we are interested in the large-system, plasmoid-dominated regime \cite[see][for review]{Loureiro_Uzdensky-2016}.  In this regime the system size is much greater than the microphysical (kinetic) plasma scales and hence where the reconnection current layer is no longer quasi-stationary and laminar, but becomes unstable and breaks up into a highly dynamic, stochastic chain of plasmoids (magnetic islands) that form a complex hierarchy.  We are especially interested in connecting the accelerated particle distribution with the statistical properties (e.g., size distribution) of the plasmoid chain. 

In its full generality, this problem, of course, is extremely complicated; after all, we are dealing here with a three-dimensional (3D), inherently kinetic (since we are interested in NTPA) system, characterized by a very large separation of length- (and hence time- and particle-energy) scales and consequently exhibiting chaotic dynamical behaviour. 
However, as mentioned above, this problem has been extensively studied in recent years with kinetic (PIC) numerical simulations, especially in two dimensions~(2D).
These studies, while computationally challenging, have produced a wealth of insightful results, essentially mapping out the NTPA quantitative characteristics --- most notably, the nonthermal particle-energy power-law index $p \equiv - {\rm d} \ln f/{\rm d}\ln \gamma$ and the high-energy cutoff $\gamma_{\rm c} m_e c^2$ --- as functions of various system parameters across several physical regimes. The most relevant parameters of the covered multi-dimensional parameter space include the hot and cold upstream plasma magnetizations ($\sigma_h$ and $\sigma_c$, respectively; see discussion in \S~\ref{subsec-picture-chain}), the guide magnetic field $B_g$, and the system size~$L$. 
Importantly, it appears that the key results from these 2D numerical studies can be summarized in terms of a few simple statements as follows.  
In the most commonly studied case of pure anti-parallel reconnection, i.e., reconnection without a guide magnetic field component, $B_g=0$, one has the following picture. First, in the ultra-relativistic reconnection limit $\sigma_h \rightarrow \infty$, the electron power-law index $p$ approaches a constant  close to and consistent with~1, independent of~$L$ if $L$ is large enough \cite[e.g.,][]{Guo_etal-2014, Werner_etal-2016}.  
The high-energy cutoff $\gamma_c$ of the nonthermal power-law segment has a nontrivial behaviour.  
For small systems, $L \lesssim L_{c} \simeq 40 \rho_0 \sigma_c$, where $\rho_0\equiv m_e c^2/e B_0$ and $B_0$ is the reconnecting upstream magnetic field, $\gamma_c$ scales linearly with system size~$L$ as $\gamma_c \sim 0.1 L/\rho _0$. This direct linear dependence simply corresponds to the available potential drop associated with the relativistic reconnection electric field $E_{\rm rec} \simeq  0.1 B_0$ and the global system size~$L$; it is sometimes called ``extreme particle acceleration" 
\cite[see][]{Aharonian_etal-2002} and corresponds to the Hillas  limit \citep{Hillas-1984}. It is, however, understood that, due to the finite energy budget, a hard spectrum with $p\leq 2$ (where the total energy budget is dominated by the most energetic particles) cannot continue to arbitrarily high energies.  
Indeed, in the large-system regime, $L>L_{c}$, the strong linear $L$-dependence of the cutoff breaks down: $\gamma_c$ rises quickly in time (roughly linearly) up to a multiple of~$\sigma_c$, e.g., up to $\gamma_{c} \simeq 4\sigma_c$ \citep{Werner_etal-2016, Kagan_etal-2018}, but then drastically slows down, and the final, asymptotic $\gamma_c$ has a  much weaker scaling with~$L$, perhaps as~$L^{1/2}$ \citep{Petropoulou_Sironi-2018, Hakobyan_etal-2020}.
Next, as the ambient $\sigma_h$ is decreased and, in particular, drops below~1, so that one enters the non-relativistic reconnection regime (the particles are still relativistic), the power-law index increases, consistent with $p \simeq C_1  + C_2 \sigma_h^{-1/2}$ \citep{Werner_etal-2018, Ball_etal-2018}, while the cutoff also decreases. 
Finally, a strong guide magnetic field $B_g$ suppresses NTPA for all~$\sigma_h$, resulting in a steeper power law and smaller~$\gamma_c$ \citep{Werner_Uzdensky-2017, Rowan_etal-2019}. 

The simplicity of these findings instills hope that it might be possible to explain them and capture the essence of 2D-reconnection-driven NTPA with a relatively simple minimal model, which would retain only the main, most critical elements of the system while perhaps neglecting various less important,  secondary details.  Constructing such a model is the main objective of this paper. 

The paper is organized as follows. 
In \S~\ref{sec-picture} we present a qualitative discussion of the basic physical picture. 
This section has two subsections: in \S~\ref{subsec-picture-chain} we describe the general properties of the plasmoid-dominated reconnection and in \S~\ref{subsec-picture-particle-acceleration} we consider how individual particles are accelerated in such a chain. 
 \S~\ref{sec-model} is devoted to the mathematical development of the proposed theory. 
 In particular, we present the general form of the kinetic equation in \S~\ref{subsec-kinetic-eqn} and then discuss its various key ingredients in subsequent subsections: the acceleration by the main reconnection electric field in~\S~\ref{subsec-accel}, particle magnetization by the inter-plasmoid reconnected magnetic field in~\S~\ref{subsec-magnetization}, particle trapping by large plasmoids in \S\S~\ref{subsec-trapping}-\ref{subsec-real-chain}, and Fermi acceleration by particle bouncing off moving plasmoids in~\S~\ref{subsec-Fermi}. We then return to the discussion of the general, full kinetic equation in~\S~\ref{subsec-general_kin_eq}.
In \S~\ref{sec-conclusions} we summarize our main findings, discuss the limitations of our present model, and outline the directions for future research.


\section{Physical Picture}
\label{sec-picture}


\subsection{Plasmoid-Dominated Reconnection Regime}
\label{subsec-picture-chain}

The main focus of this paper is on nonthermal particle acceleration. Efficient high-energy NTPA requires strong electric fields coherent over substantial distances, at least comparable or larger than the Laromor radii of energetic particles and therefore much larger than the plasma kinetic microscales. On such scales the electric field should be the motional (${\bf u\times B}$) ideal-MHD electric fieldd \cite[c.f., e.g.,][]{Drake_etal-2019}. This simple reasoning underlies the need to understand the origin and structure of bulk plasma motions. For the resulting electric fields to be strong, these motions need to be fast, e.g., Alfv\'enic. Often, including in the plasmoid-dominated magnetic reconnection regime, the relevant fast motions arise as a result of the nonlinear development of various instabilities. Thus, it is important to identify what instabilities operate in a reconnecting current layer (or a reconnecting plasmoid chain) and what motions they drive. We will argue that 2D plasmoid-mediated reconnection layers are subject to two important instabilities: the secondary tearing (aka plasmoid) instability, leading to the growth of plasmoids, and the coalescence instability, causing the plasmoids to move towards or away from each other along the global layer. These two instabilities lead to bulk plasma motions along the layer, and the associated electric fields lead to two channels for~NTPA, as we will explain below.

In this section, we outline the basic physical picture of plasmoid-dominated reconnection that we believe adequately captures the physics most relevant for NTPA. This underlying picture is qualitatively the same for any plasma-physical framework of 2D reconnection: resistive or collisionless, relativistic or non-relativistic. 

For simplicity, we will consider only 2D reconnection, thus ignoring 3D effects. In addition to the simplicity considerations, this approximation is motivated by the observation that recent 3D PIC studies (e.g., \citealt{Werner_Uzdensky-2017}; see also \citealt{Sironi_Spitkovsky-2014}) indicate that 3D collisionless magnetic reconnection produces nonthermal particle spectra that are quite similar to the spectra produced by its 2D counterpart, at least for relativistic reconnection in pair plasmas (see, however, \citealt{Dahlin_etal-2015, Dahlin_etal-2017} for 2D/3D comparison studies of NTPA in nonrelativistic reconnection with a finite guide field, and Werner \& Uzdensky 2020, {\it in preparation}, for trans-relativistic reconnection), even though the layer's morphology may be quite different. 
In addition, most of the numerical PIC studies of magnetic reconnection have so far been done only in~2D and hence most of what we know about reconnection-driven relativistic NTPA is limited to the 2D case; it therefore makes sense to focus on the 2D case first, before tackling the general 3D situation. 
We acknowledge, however, that the role of the system dimensionality remains an open issue and should be investigated further.%
\footnote{We also limit the present paper to the case of non-radiative magnetic reconnection. We note, however, that the case of radiative reconnection, where radiation back-reaction on the emitting particles has a strong influence on high-energy NTPA and may even affect the general dynamics and energetics of the reconnection process is a vibrant and rapidly developing area of current research with strong astrophysical motivation, and we refer the interested reader to the recent papers on this subject \citep{Jaroschek_Hoshino-2009, Uzdensky-2011, Uzdensky_McKinney-2011, Uzdensky_etal-2011, Nalewajko_etal-2011, Nalewajko_etal-2012, McKinney_Uzdensky-2012, Cerutti_etal-2012a, Cerutti_etal-2013, Cerutti_etal-2014a, Beloborodov-2017, Nalewajko_etal-2018, Werner_etal-2019, Schoeffler_etal-2019, Hakobyan_etal-2019, Sironi_Beloborodov-2019, Ortuno-Macias_Nalewajko-2019, Mehlhaff_etal-2020}, see \cite{Uzdensky-2016} for a review.}

\begin{figure}
\begin{center}
\includegraphics[width=10 cm]{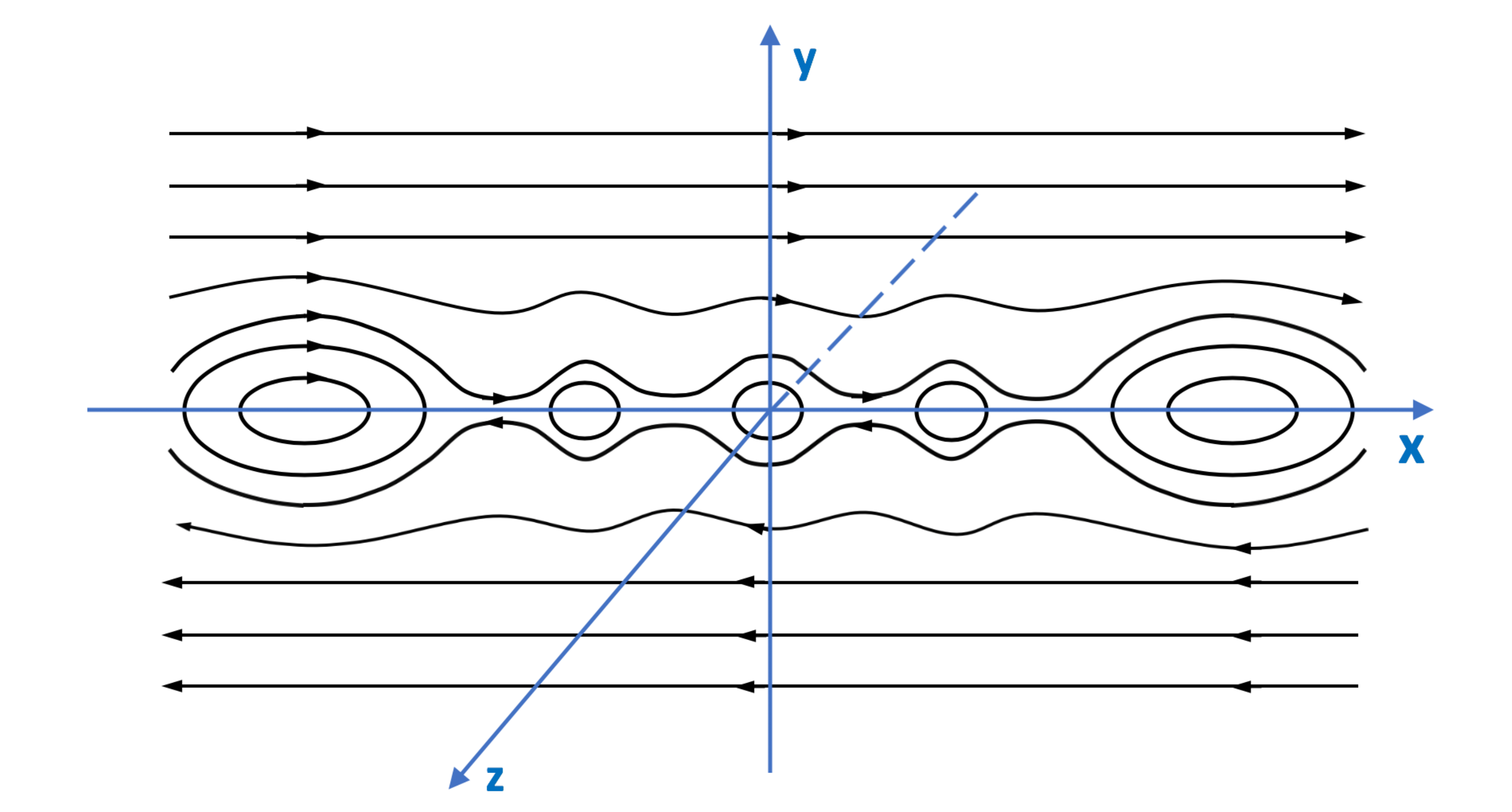}
\end{center}
\caption{The coordinate system (blue) used in this article. The black lines show magnetic field lines in a plasmoid-dominated reconnection layer. The $x$-direction is the direction of the reversing reconnecting magnetic field, the $y$-direction is across across the current layer (the direction of the field reversal), and the $z$-direction is that of the main electric field and current in the sheet (the ignorable direction).}
\label{fig-1}
\end{figure}

For convenience, we introduce the following system of coordinates (see figure~\ref{fig-1}): 
$x$ is the direction of the reversing reconnecting magnetic field ${\bf B_0}$ (the outflow direction); 
$y$ is the direction perpendicular to the current layer (the inflow direction); 
and $z$ is the direction of the electric current and of the main reconnection electric field $E_{\rm rec}$ (sometimes called the out-of-plane, or ignorable direction). 
In general, there may also be a guide magnetic field $B_g$ in the $z$ direction. 
Together, $x$ and $y$ form what is often called the ``reconnection plane", and $x$ and $z$ form the reconnection-layer midplane (at $y=0$).


We envision a vigorous reconnection process taking place in the large-system, plasmoid-dominated regime \cite[e.g.,][]{Shibata_Tanuma-2001, Bhattacharjee_etal-2009, Daughton_etal-2009, Uzdensky_etal-2010, Huang_Bhattacharjee-2010, Loureiro_etal-2012, Sironi_etal-2016, Werner_etal-2018, Werner_Uzdensky-2017, Petropoulou_etal-2016, Petropoulou_etal-2018, Schoeffler_etal-2019}. 
The reconnection layer is broken up into a highly dynamic hierarchical plasmoid chain as a result of the secondary-tearing, aka plasmoid, instability \citep{Loureiro_etal-2007, Daughton_etal-2009, Samtaney_etal-2009, Bhattacharjee_etal-2009, Uzdensky_etal-2010, Loureiro_etal-2012, Loureiro_etal-2013}; (see \citealt{Loureiro_Uzdensky-2016} for a recent review).  
The chain is essentially one-dimensional (1D), consisting of a broad distribution of plasmoids (aka magnetic islands, centered around magnetic O-points) of different sizes, strung on a single line $y=0$  (which is a projection of the reconnection midplane) and connected to each other by reconnecting inter-plasmoid current layers containing magnetic X-points (see figure~\ref{fig-1}).  The plasmoids grow (i.e., accumulate magnetic flux and mass) continuously via reconnection taking place in these inter-plasmoid layers. The $E_z$ electric field associated with these inter-plasmoid reconnection processes would exist even if the plasmoids were themselves stationary; this field is responsible for one of the channels of particle acceleration and thus will play an important role in our analysis, as discussed in \S~\ref{subsec-picture-particle-acceleration}.

In reality, however, the plasmoids are not stationary --- they move about in the reconnection midplane (i.e., in the $x$ direction) with different velocities (generally, of order the Alfv\'en speed~$V_A$) in a complicated, chaotic fashion, a kind of 1D plasmoid turbulence.  Sometimes they merge with each other, overall maintaining a broad statistical distribution of sizes (see \S\S~\ref{subsec-plasmoid_chain-simple}-\ref{subsec-real-chain} for a more detailed description).  
The flow dynamics that controls plasmoid motions in the $x$-direction is complex and is governed by the interplay of two factors: the coalescence instability, developing on a broad range of scales, and the large-scale inhomogeneities along the global layer, which drive divergent large-scale flows (Hubble flow, $u_x \sim x$) out of the layer. 
Overall, one may distinguish between two somewhat different situations, which we may call (1) the nonlinear evolution of a tearing-unstable plasmoid chain and (2) reconnection proper; this distinction is somewhat similar in spirit to that made in turbulence studies between, respectively, decaying and driven turbulence. 

In the first situation, which can be appropriately called the nonlinear evolution of a plasmoid chain and which represents a standard choice for numerical studies of reconnection (e.g., \citealt{Zenitani_Hoshino-2001, Zenitani_Hoshino-2005, Zenitani_Hoshino-2008, Jaroschek_etal-2004a, Cerutti_etal-2012b, Cerutti_etal-2013, Guo_etal-2014, Guo_etal-2015, Werner_etal-2016, Werner_Uzdensky-2017, Werner_etal-2018}), one imposes periodic boundary conditions in the $x$-direction. One then considers the evolution of a preexisting current sheet, usually taken to be rather thin.  The current sheet is often set up initially to be translationally symmetric in the $x$-direction or may have a small initial magnetic perturbation (often sinusoidal in~$x$) imposed to trigger the onset of tearing faster. 
In either case, a true statistical steady state is not possible and, instead, the system undergoes a complicated evolution, which, however, has well-defined initial and final states. The long and thin initial current layer becomes unstable to the tearing instability and quickly breaks up into a multitude of primary, 1st-generation plasmoids, corresponding to the fastest growing tearing mode \citep{Samtaney_etal-2009}, thus breaking the translational symmetry in the $x$-direction. As these plasmoids grow, first linearly and then nonlinearly, at some point they start feeling and interacting with each other. 
Namely, the still-growing plasmoid chain becomes subject to the coalescence instability \citep{Finn_Kaw-1977, Jaroschek_etal-2004a, Daughton_Karimabadi-2007, Oka_etal-2010} that makes a given plasmoid decide to move towards its neighbor either on the left or on the right, thus causing the plasmoids to pair up and merge (coalesce) with their neighbors. 
Since the coalescence instability requires the existence of plasmoids in the first place, and since its growth rate increases as the plasmoids grow, it can be regarded as a secondary, or parasitic, instability with respect to the primary tearing mode. The coalescence instability initiates plasmoid motions along the $x$-axis, which are, however, constrained by the imposed periodic boundary conditions. 
As these primary plasmoids merge with each other hierarchically, their number decreases and the distance between them increases. Correspondingly, the secondary inter-plasmoid current sheets between these plasmoids get stretched and can themselves become tearing-unstable, resulting in the production of the next generation of secondary plasmoids, and so on, establishing a hierarchical structure. 
Importantly, any large plasmoid perturbs the magnetic field, and hence the magnetic forces and the flow structure, around it, serving as an attractor for nearby small plasmoids; thus, the secondary current layer (which, in general, is itself a hierarchical plasmoid subchain) between any two large plasmoids develops an internal relative stagnation point directing the plasma outflows towards the two large plasmoids. 
In addition, when two large plasmoids merge with each other, they do this via reconnection taking place in a secondary current sheet that is perpendicular to the main one, and, if the merging plasmoids are large enough, may itself become a reconnecting secondary plasmoid chain \citep{Daughton_Karimabadi-2007, Pritchett-2008, Oka_etal-2010}. 
All these interacting and concurrent processes result in an nontrivial and complicated intermediate nonlinear dynamical stage that we can call the active reconnection phase. The system, however, is continuously evolving even in the statistical sense and does not have a clear long-term actively-reconnecting statistical steady state. 
Eventually, the largest plasmoids grow so big that their size in the $y$ direction becomes comparable to the $x$-separation between them, forcing the current sheets between them to shrink back into nearly-$90^\circ$ X-point configurations. This causes reconnection to stop: the tearing mode saturates and the plasmoids stop growing. The system then approaches the final relaxed state characterized by a small number of remaining big plasmoids, dictated essentially by the aspect ratio of the box~$L_y/L_x$:  if $L_y \gtrsim L_x$, one has just one plasmoid (and hence just one magnetic O-point and one X-point); but in the case of an elongated box, $L_y \ll L_x$, one may have a stable chain of many plasmoids \cite[e.g.,][]{Jaroschek_etal-2004a}.  The time evolution of the system in either case thus has a well-defined end state, which can be used for characterizing the properties and overall effectiveness of NTPA in this scenario.

The situation is somewhat different, however, in the second situation, which we will call reconnection proper. In this case, instead of periodic boundary conditions, one uses free outflow boundary conditions in the $x$-direction \citep{Daughton_etal-2006, Loureiro_etal-2012, Sironi_Spitkovsky-2014} which, however, are somewhat difficult to implement in PIC~\citep{Daughton_etal-2006, Daughton_Karimabadi-2007}. These outflow boundary conditions in the $x$-direction are usually supplemented with free inflow boundary conditions at the $y$-boundaries or an indefinitely expanding box in the $y$-direction  \citep{Sironi_Spitkovsky-2014}. 
In this case, generically, there is a dominant global stagnation point (coinciding with the main magnetic X-point) that plays the role of a continental divide that separates (causes a bifurcation of) the overall, large-scale reconnection outflows from the layer.  These  outflows are sheared, i.e., $u_x \sim x$ (like the Hubble flow), as the flow accelerates along the sheet; this longitudinal shear may stabilize the tearing instability somewhat (e.g., \citealt{Bulanov_etal-1978, Loureiro_etal-2007, Loureiro_etal-2013}), but cannot suppress it completely for sufficiently large systems~\citep{Tolman_etal-2018}. 
One then has all the complex intermediate-stage dynamics of the nonlinear tearing chain described above in the previous paragraph, plus the overall large-scale outward motion and eventual ejection of plasmoids out of the layer~\citep{Uzdensky_etal-2010}. 
Because the large-scale flow is sheared, the inter-plasmoid current layers continuously get stretched and become tearing-unstable, giving birth to new, next-generation plasmoids. 
The lifetime of an individual plasmoid is limited by the time before it gets swallowed by a bigger plasmoid or is ejected out of the system \citep{Uzdensky_etal-2010, Loureiro_etal-2012, Sironi_etal-2016}, the latter typically of the order of the Alfv\'en crossing time along the layer (although it may be longer by a logarithmic factor of a few for so-called ``monster plasmoids" born very close to the global main X-point, see \citealt{Uzdensky_etal-2010}). 
As plasmoids continuously appear, grow, merge, move out, get ejected, and are replaced by new plasmoids, the system may exist indefinitely in a statistical steady state \citep{Loureiro_etal-2012, Sironi_etal-2016}, which makes this configuration attractive for numerical studies. 

In any case, we envision that the hierarchical plasmoid chain during the active reconnection stage has a self-similar, fractal structure (see figure~\ref{fig-2}), as was first proposed by \cite{Shibata_Tanuma-2001} and then confirmed numerically in both the resistive-MHD regime \citep{Bhattacharjee_etal-2009, Huang_Bhattacharjee-2010, Loureiro_etal-2012} and collisionless regime \citep{Daughton_etal-2009, Daughton_etal-2011, Sironi_etal-2016, Werner_etal-2016, Werner_etal-2018}. This means that, when one looks closely at an inter-plasmoid current layer between two neighboring plasmoids of similar size $w$ at some given level in the middle of the hierarchy, one again finds a plasmoid-dominated reconnection region with essentially the same, universal characteristic values of three key electromagnetic (EM) field components as found at all other levels in the hierarchy. These field components are: the reconnecting ($x$-direction) upstream magnetic field~$B_0$ (and hence the corresponding Alfv\'en speed~$V_A$, which sets the characteristic scale for the plasma motions along the layer), the inter-plasmoid reconnected (i.e., in the $y$ direction) magnetic field~$B_1$ (e.g., averaged over one half of the inter-plasmoid layer under consideration), and the effective reconnection rate~$E_{\rm rec}$ (i.e., the electric field in the $z$ direction). This self-similarity extends all the way from the global layer as a whole (of size~$L$) at the top of the hierarchy down to the smallest {\it elementary current layers} (which are marginally stable to tearing and thus essentially laminar) at the very bottom of the hierarchy, of characteristic thickness~$\delta$ and length~$\ell$.  In the case of collisionless reconnection without a strong guide field, $\delta \sim \bar{\rho}$ (the typical Larmor radius of particles in the layer in the upstream reconnecting magnetic field~$B_0$) and $\ell \sim 10-30\, \delta$. At these smallest scales, the self-similarity of the EM field structure breaks down; in particular, in the case of electron-ion plasma reconnection, the Hall effect becomes important at these scales, leading to the emergence of an quadrupole out-of-plane ($z$-direction) magnetic field and an in-plane bipolar electrostatic electric field~\citep{Sonnerup-1979, Terasawa-1983, Shay_etal-1998, Uzdensky_Kulsrud-2006, Melzani_etal-2014a, Werner_etal-2018}. We will ignore these fields in the present analysis because they only affect the typical, average-energy particles, but not the highly energetic, nonthermal particles of interest to us here. 

All three above-mentioned field quantities --- $B_0$, $B_1$, and $E_{\rm rec}$ --- will play important roles in our analysis. 
Since we are ultimately interested in the effects that these fields have on NTPA, it is important to recognize that what matters is how the structure of these fields appears to a given energetic 
(i.e., with energy $\gamma m_e c^2$ well above the average particle energy in the layer, $\bar{\gamma} m_e c^2$) 
particle under consideration. Such a particle will have a large Larmor radius $\rho(\gamma)$ and so its motion will be blind, nearly insensitive to small-scale EM structures of modest field strength. It will interact most strongly with EM structures that exist on scales of order its Larmor radius or larger.  It is thus important to look at the EM fields as a function of scale $\lambda$ (in the $x$-direction).  This philosophy is similar to the one used in analyzing the self-similar dynamics in the inertial range of turbulence (on scales  much smaller than the large driving scale but much larger than the dissipative scale) and, more specifically, turbulent NTPA, where one is interested primarily in the resonant interaction of a particle with turbulent eddies or waves of a given size. 

While $B_0$ is relatively straightforward, the other two quantities ($B_1$ and~$E_{\rm rec}$) are less trivial and deserve a careful discussion. We first discuss the nature of the reconnected magnetic field $B_1$ (in the $y$-direction) in the plasmoid-dominated reconnection regime. 
Let us consider it somewhere in the middle of the plasmoid hierarchy, i.e., on the scale of an inter-plasmoid reconnecting layer (or, more precisely, inter-plasmoid reconnecting chain) between two plasmoids of some intermediate size~$w$ such that $\delta \ll w \ll L$. 
The length of this inter-plasmoid current layer, $\lambda$, is then much smaller than the global length of the layer~$L$, but larger than the length $\ell$ of the shortest elementary inter-plasmoid current sheets; in simulations it is typically $\sim$10 times greater that the width $w$ of the two plasmoids flanking it.  
The reconnected magnetic field~$B_1$ can be defined, e.g., in terms of the net reconnected ($B_y$) magnetic flux between the layer's X point and the edge of adjacent plasmoid, divided by~$\lambda/2$. 
While there may be many smaller-size (i.e., belonging to the next levels of the plasmoid hierarchy) magnetic islands inhabiting the inter-plasmoid layer under consideration, they consist of closed magnetic flux surfaces and thus do not contribute to the $\lambda$-scale~$B_1$. Instead, this field is build up from patches of ``semi-open" reconnected flux --- the field lines that intersect the reconnection midplane only once in the region between the two big ($w$-scale) islands under consideration; these field lines may then envelope these plasmoids and close (i.e., intersect the reconnection midplane again) somewhere outside the inter-plasmoid layer in question (see \citealt{Uzdensky_etal-2010}).  The self-similarity of the plasmoid chain then dictates that the typical value of $B_1$ is independent of the scale within the hierarchy. It is thus comparable to the typical reconnected magnetic field values in the outflow exhaust regions of elementary current layers at the very bottom of the hierarchy, i.e., about $\epsilon B_0 \simeq 0.1\,B_0$ for collisionless reconnection (and about 0.01\,$B_0$ for collisional, resistive~MHD reconnection; but in this study for concreteness we will adopt the fiducial collisionless value of~0.1\,$B_0$). 
This point will be important in our analysis when we consider the motion of energetic particles in such a layer. 

Likewise, the self-similar reconnecting plasmoid chain is characterized by a universal (scale-independent) typical average value of the out-of-plane electric field~$|E_z| = E_{\rm rec}$ in the inter-plasmoid current sheet at all levels of the hierarchy. This {\it effective reconnection rate} is thus equal to the characteristic {\it microscopic reconnection rate} at individual inter-plasmoid X-points in the elementary current layers \citep{Uzdensky_etal-2010}. This electric field is associated with the nonlinear development of the secondary tearing instability. 
It can be conveniently parametrized in terms of the upstream asymptotic magnetic field $B_0$ as 
\beq
E_{\rm rec} = v_{\rm rec} B_0/c = \beta_{\rm rec} B_0 = \epsilon B_0 V_A/c \, , 
\label{eq-E_rec}
\eeq
where the dimensionless coefficients $\beta_{\rm rec} \equiv v_{\rm rec}/c$ and $\epsilon = v_{\rm rec}/V_A$ represent the reconnection inflow speed $v_{\rm rec}$ normalized, respectively, to the speed of light~$c$ and to the Alfv\'en speed $V_A$ defined with the upstream reconnecting field~$B_0$ and discussed in more detail below.
First-principles PIC simulations  \cite[e.g.,][]{Hesse_etal-1999, Birn_etal-2001, Werner_etal-2018} and analytical theories \citep{Comisso_Bhattacharjee-2016, Cassak_etal-2017} indicate that  the dimensionless reconnection rate $\epsilon$ for collisionless reconnection is typically $\epsilon \simeq 0.1$ in both relativistic and non-relativistic regimes, the value that we will adopt in this paper.%
\footnote{For reference, the dimensionless reconnection rate $\epsilon$ is about 0.01 in plasmoid-dominated resistive MHD reconnection \citep{Bhattacharjee_etal-2009, Uzdensky_etal-2010, Huang_Bhattacharjee-2010, Loureiro_etal-2012}.} 
This is true even for electron-positron pair plasmas (\citealt{Bessho_Bhattacharjee-2005, Bessho_Bhattacharjee-2007}), despite the absence of the Hall effect that has been linked to fast ($\epsilon \simeq  0.1$) collisionless reconnection in electron-ion plasmas in previous studies \cite[e.g.,][]{Birn_etal-2001}.  

In a laminar and stationary 2D reconnection layer, the electric field $E_z$ would be steady and uniform, as dictated by Faraday's law. 
In contrast, however, a realistic stochastic reconnecting plasmoid chain is highly dynamic and $E_z$ can vary dramatically in both space and time due to rapid (Alfv\'enic) chaotic plasmoid motions up and down the chain, e.g., driven by the plasmoid coalescence instability as discussed above. 
The electric field at any given point can then have strong fluctuations on top of the average value of 0.1~$V_A B_0/c$. These fluctuations may have the form of rapid, intense alternating-sign spikes of the motional electric field of amplitude as high as $E_{\rm pl} = V_A B_0/c$ (i.e., 10 times the mean), e.g., when a circularized plasmoid with $B_y \sim B_0$ passes by with $v_x \sim V_A$ \cite[e.g.,][]{Zenitani_Hoshino-2001, Zenitani_Hoshino-2005, Sironi_etal-2016, Philippov_etal-2019}.
It is this strong motional electric field that is responsible for short intense particle acceleration events associated with an energetic particle bouncing off a rapidly moving plasmoid --- a key element in what is described as stochastic Fermi acceleration.%
\footnote{In principle, the motional electric field may sometimes even exceed~$V_A B_0/c$; indeed, in the absence of a strong guide field, the circular magnetic field deep inside a large plasmoid's core is squeezed by the pinch force to values much higher than~$B_0$, especially when radiative cooling is strong \cite[e.g.,][]{Schoeffler_etal-2019, Werner_etal-2019}; when such a plasmoid moves with $v_x \sim V_A$, the local motional $E_z = v_x B_y/c$ can be larger than~$V_A B_0/c$.  
Furthermore, when two plasmoids approach each other and start merging, the so-called anti-reconnection electric field in the perpendicular secondary reconnecting current sheet between them is reversed relative to $E_z$ in the main layer. In this paper, however, we will ignore these complications, leaving them for future study.}

As will be described in more detail in \S~\ref{subsec-picture-particle-acceleration}, the model presented in this paper will incorporate both acceleration mechanisms, viewed as separate channels: acceleration by $E_{\rm rec}$, as a particle traverses an inter-plasmoid current layer (the motional component of this electric field is associated with the reconnection outflows out of the individual inter-plasmoid current sheets); and acceleration by $E_{\rm pl}$ as it bounces off a large rapidly moving plasmoid. 
An essential assumption that is important to us here is that the statistical properties of the electric field, e.g., the distribution of the electric field strengths and the average value, be the same at all level of the hierarchy, i.e., for reconnecting sublayers of each scale.  


The relativistic Alfv\'en speed $V_A$ that appears in equation~(\ref{eq-E_rec}) is associated with the reconnecting magnetic field $B_0$ and is defined in terms of the ambient upstream plasma conditions. In particular, it is convenient to express it in a dimensionless form 
\beq
V_A \equiv \beta_A c = c \, \sqrt{\sigma_h \over{1+\sigma_h}} 
\label{eq-V_A}
\eeq
in terms of the so-called ``hot" upstream magnetization parameter $\sigma_h$, defined as the ratio of the enthalpy density of the reconnecting magnetic field $B_0$ to the relativistic (including rest-mass) enthalpy density $h$ of the upstream plasma \citep{Melzani_etal-2014a, Werner_etal-2018},
\beq
\sigma_h \equiv {{B_0^2}\over{4 \pi h}} \, . 
\label{eq-sigma_h-def}
\eeq

For example, in the case of a pair plasma that is relativistically cold, i.e., has an upstream background temperature $T_b = \theta_e m_e c^2 \ll m_e c^2$, or in the case of a pure electron-ion plasma (with $n_{e,b} = n_{i,b}$) that is nonrelativistic ($T_b \ll m_e c^2$) or semi-relativistic ($m_e c^2 \ll T_b \ll m_i c^2$), the enthalpy is dominated by the rest-mass of the dominant particle species (electrons and positrons in the pair plasma case and ions in the electron-ion plasma case).  In these cases, the ``hot" upstream magnetization becomes the same as the ``cold" plasma magnetization for this dominant species:
\beq
\sigma_{c} \equiv {{B_0^2}\over{4 \pi n_{b} m_e c^2}} 
\label{eq-sigma_c-def-pair}
\eeq
for the pair case with $n_b = 2 n_{b,e}$ being the total (electron + positron) particle number density, 
and 
\beq
\sigma_{c} \simeq \sigma_{ci} \equiv  {{B_0^2}\over{4 \pi n_{ib} m_i c^2}} \, , 
\label{eq-sigma_c-def-ei}
\eeq
for the electron-ion case.

However, in the opposite case of an upstream plasma that is ultra-relativistically hot, i.e., $T_b = \theta_e m_e c^2 \gg m_e c^2$  for the pair plasma case (as found in, e.g., PWN) or $T_b \gg m_i c^2$  for the electron-ion plasma case, the rest-mass contribution to the upstream enthalpy is negligible and so the enthalpy density becomes simply 
$h \simeq 4 P_b = 4 n_b T_b = 4 n_b \theta_e m_e c^2 $ (where $n_b$ is the total background particle density). 
In this case, the ``hot" magnetization simply becomes (apart from a factor of 1/2) the inverse of the upstream plasma-$\beta$ parameter 
($\beta_b \equiv 8\pi P_b/B_0^2$),  
\beq
\sigma_h(T_b \gg m_e c^2, m_i c^2)  = {{B_0^2}\over{16 \pi n_b T_b}} =  {1\over{2\beta_b}}, 
\eeq
and thus differs dramatically from the ``cold" magnetization, e.g., $\sigma_h = \sigma_c/4 \theta_e \ll \sigma_c$ for the pair-plasma case and   $\sigma_h = \sigma_{ci}/8 \theta_i \ll \sigma_{ci}$ for the electron-ion case with $n_{i,b}=n_{e,b}$. 

Both $\sigma_h$ and $\sigma_c$ are useful quantities that will play important roles in our analysis. 
In particular, $\sigma_c$ reflects (up to a factor or order unity) the upstream magnetic energy per background particle and thus sets the basic characteristic energy scale for particles energized by the reconnection process (normalized by the particle rest mass). 
And the hot magnetization $\sigma_h$ controls how relativistic the upstream Alfv\'en velocity---and hence the bulk fluid motions in the layer---are. Namely, it allows us to distinguish two limiting cases: \\
- the {\it relativistic reconnection regime}:  
$\sigma_h \gg 1$ and hence $\beta_A \equiv V_A/c = [\sigma/(1+\sigma)]^{1/2} \rightarrow 1$ and $E_{\rm rec}  \simeq \epsilon B_0 \simeq 0.1 B_0$;  \\
- the {\it nonrelativistic reconnection regime}: 
$\sigma_h \ll 1$ and hence $\beta_A \simeq \sigma_h^{1/2} \ll 1$ and $E_{\rm rec}  \simeq \epsilon \sigma_h^{1/2} B_0$.

For the sake of completeness, we mention here how some of the above key relationships are modified in the presence of a guide ($\hat{z}$-component) magnetic field~$B_g$, even though this paper focusses on the zero-$B_g$ case. Following \cite{Werner_Uzdensky-2017}, in the case of a finite guide field, the plasma outflows that control the reconnection electric field correspond not to the full Alfv\'en velocity but  to its projection onto the outflow ($x$) direction, i.e., to the in-plane Alfv\'en velocity:
\beq
{{V_{A,x}^2}\over{c^2}} = {B_0^2\over{B_{\rm tot}^2}} \, {{B_{\rm tot}^2}\over{B_{\rm tot}^2 + 4 \pi h}}  = 
{{B_0^2}\over{B_{\rm tot}^2 + 4 \pi h}} \, , 
\eeq
where $B_{\rm tot}^2 = B_0^2 + B_g^2$. 
This expression can be rewritten as 
\beq
{{V_{A,x}^2}\over{c^2}} = {{B_0^2}\over{B_0^2 + (B_g^2+ 4 \pi h)}}  = 
{{B_0^2}\over{B_0^2 + 4 \pi h_{\rm eff}}}  \, , 
\eeq
where we have introduced the effective total enthalpy, $h_{\rm eff} \equiv h + B_g^2/4\pi$, that includes the contribution of the guide magnetic field, $B_g^2/4\pi$, in addition to the relativistic plasma enthalpy~$h$. 
Physically, this additional contribution needs to be included because the guide field is advected together with the plasma out of the layer by the (Alfv\'enic) reconnection outflows, and this its inertia (and, more specifically, its enthalpy) has to be taken into account \citep{Werner_Uzdensky-2017}.  With this in mind, we can write the relevant in-plane Alfv\'en velocity in the standard form: 
\beq
{{V_{A,x}^2}\over{c^2}} = {\sigma_{\rm eff} \over{1+\sigma_{\rm eff}}} \, , 
\eeq
where we have defined the {\it effective hot magnetization} \citep{Werner_Uzdensky-2017}
\beq
\sigma_{\rm eff} \equiv {{B_0^2}\over{4 \pi h_{\rm eff}}} = {{B_0^2}\over{(B_g^2+ 4 \pi h)}} \, . 
\eeq

Thus, in the case of a relativistically-strong guide field, $B_g^2 \gg 4 \pi h$, when the plasma enthaply is negligible and we are dealing with a relativistic force-free field, we get $\sigma_{\rm eff} \rightarrow B_0^2/B_g^2$, and so $V_{A,x}/c \rightarrow B_0/B_{\rm tot}$.  And in the opposite, non-relativistic case, $B_g^2 \ll 4 \pi h$, (note that this does not imply that the guide field is weak compared to the reconnecting field~$B_0$), the effect of the guide field on $\sigma_{\rm eff}$ and $V_{A,x}$ can be ignored and we recover the standard expression $V_{A,x} = c\, [\sigma_h/(1+\sigma_h)]^{1/2}$.


\subsection{Acceleration of Relativistic Particles in a Reconnecting Plasmoid Chain}
\label{subsec-picture-particle-acceleration}

We shall now discuss the motion and acceleration of energetic relativistic particles in the reconnecting plasmoid chain. 
We will focus here on relativistic particles with energies $\gamma m_e c^2$ in the high-energy nonthermal  tail of the distribution function, far above the average particle energy~$\bar{\gamma} m_e c^2$. We will call such particles simply ``energetic particles" or ``high-energy particles". 
The Larmor radii of such particles (corresponding to the upstream magnetic field $B_0$), given by $\rho(\gamma) = \gamma \rho_0$, where $\rho_0 \equiv m_e c^2/e B_0$, are  much greater than the average electron Larmor radius, $\bar{\rho} \equiv \bar{\gamma} \rho_0$.  Since the thickness $\delta$ of the smallest elementary current layers in collisionless reconnection is usually of order $\bar{\rho}$, this means that the energetic particles under consideration will have Larmor radii greater than~$\delta$.  

Because of this, the questions of whether the acceleration of such highly energetic particles is done by a nonideal electric field or by an ideal (motional) electric field, or whether this accelerating electric field is mostly parallel or perpendicular to the local magnetic field, may not be very relevant or even well posed.  A given highly energetic particle does not know or care whether the electric field accelerating it is ideal or nonideal. What matters is just the electric field component along the particle's direction of motion, smoothed on the appropriate scale of the particle's motion. For a particle with $\gamma \gg \bar{\gamma}$, this scale is generally larger than the microscopic plasma scales (like $\delta$ or $\bar{\rho}$), on which one can determine whether the electric field is ideal or not. Thus, the electric field responsible for energetic particle acceleration may in general have both ideal and nonideal components, but in our view this question is not particularly relevant for understanding NTPA of highly energetic particles.%
\footnote{This issue may, however, still be important for analyzing the energization of just slightly supra-thermal particles, which governs the particle injection into the nonthermal tail.}
Moreover, the distinction between ideal and nonideal electric fields may be even less relevant in the kinetic picture; this concept requires identifying the plasma bulk velocity ${\bf u}$, i.e., the average particle velocity, but individual particles will have velocities that may in general be very different from~${\bf u}$. Furthermore, in relativistic plasmas even defining the plasma's bulk velocity is not a completely trivial task, as there are two different ways of doing this [the \cite{Landau_Lifshitz-1959} and the \cite{Eckart-1940} frames]. In this case, it may be sensible to define the nonideal parallel electric field in the de Hoffmann--Teller frame, which corresponds to the ${\bf E\times B}$ drift  and in which the perpendicular (to the magnetic field) electric field component vanishes. 

Let us now discuss the main characteristics of an accelerating energetic particle's motion. 
The first two key features of  this motion are: 
{\it (i)} confinement to the reconnection layer in the $y$ direction (i.e., across the layer) by the reversing upstream magnetic field, which always deflects the particle back towards the layer; 
and 
{\it (ii)} the particle's continuous and relatively steady acceleration along the layer in the $z$ direction by the main reconnection electric field~$E_{\rm rec}$; along with the electric field $E_{\rm pl}$ involved in the Fermi acceleration by particle reflection off of moving large plasmoids (discussed below), it is this electric field that is ultimately responsible for primary particle acceleration in our model. 
Thus, the predominant motion of the particle in the $yz$ plane during its acceleration stage (i.e., until it gets magnetized by the reconnecting ($B_y$) field and eventually trapped by a large plasmoid, see below) can be described by the relativistic version of a Speiser orbit \citep{Speiser-1965, Zenitani_Hoshino-2001, Uzdensky_etal-2011, Cerutti_etal-2012a}, where the particle wiggles in and out of the thin current layer into the region of stronger ($B_0$) upstream magnetic field, crossing the reconnection midplane $y=0$ multiple times as it continuously gains energy. Since the particle's Larmor radius in the upstream magnetic field is greater than~$\delta$, as we discussed above, the particle may initially spend a substantial fraction of time in one of the two upstream regions on either side of the current sheet. However, as the particle moves along this trajectory and is accelerated by the reconnection electric field, over time its relativistic Speiser trajectory focusses closer and closer to the midplane, with the $y$-meandering getting progressively smaller \citep{Kirk-2004, Contopoulos-2007b, Uzdensky_etal-2011, Cerutti_etal-2012a}, and hence the particle's confinement to the layer becoming tighter. 

The most interesting and nontrivial dynamics thus takes place in the $x$ direction, i.e., along the reconnecting magnetic field.  
The main field component controlling this motion is the reconnected magnetic field~$B_y$. The particle's interaction with~$B_y$ governs the lifetime of the particle in the active acceleration region before it effectively escapes from it (see below); it involves its deflection away from X-points by the distributed reconnected magnetic field $B_1$ and the particle's interaction with sufficiently large magnetic islands (plasmoids). 

\begin{figure}
\begin{center}
\includegraphics[width=5.cm]{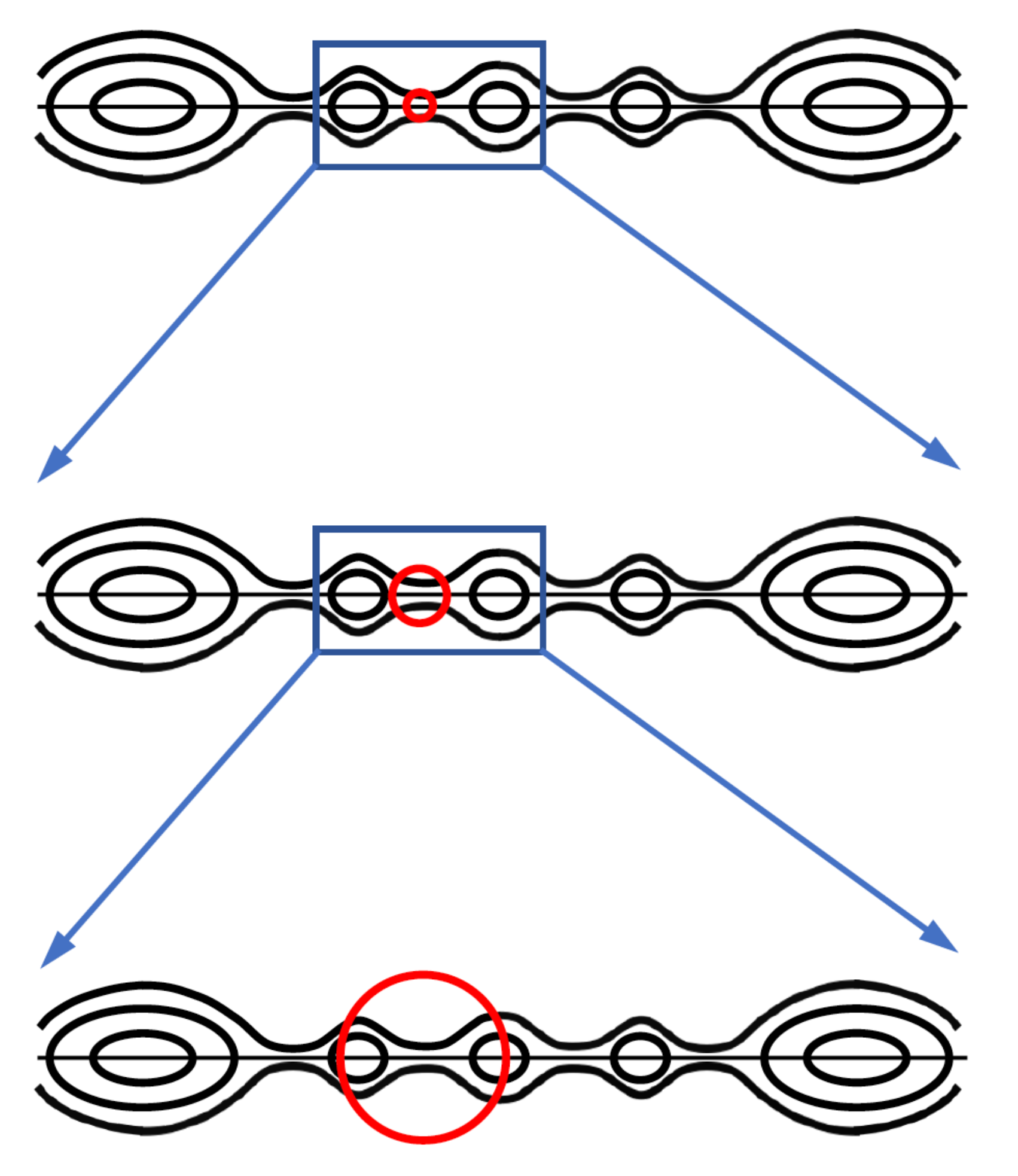}
\end{center}
\caption{Characteristic Larmor radius of a given energetic particle (red circle) appears differently relative to the sizes of the neighboring plasmoids in a self-similar hierarchical (Shibata-Tanuma) plasmoid chain (black), when viewed at different levels in the hierarchy. It is smaller than its flanking plasmoid sizes at the top level, comparable to them at the middle level, and bigger at the bottom level.}
\label{fig-2}
\end{figure}

We stress that the motion of an energetic particle of a given energy~$\gamma m_e c^2$ in the $x$-direction through a hierarchical plasmoid chain should be analyzed at the appropriate level of the plasmoid hierarchy (see figure~\ref{fig-2}), corresponding to intermediate-scale plasmoid-mediated reconnection layers between adjacent plasmoids of size $w(\gamma)  \sim \rho_L(\gamma)$, large enough to reflect of confine the particle. To this particle, the space between two such large plasmoids effectively looks like a reconnecting current layer of length corresponding to the typical inter-plasmoid separation~$\lambda_{\rm pl}[w(\gamma)]$, with a reconnection electric field $E_{\rm rec}$ and a typical reconnected magnetic field~$B_1$, as discussed in \S~\ref{subsec-picture-chain} above.
The fact that this layer is not a small elementary layer at the very bottom of the plasmoid hierarchy, but instead is somewhere in the middle of it and hence itself has a nontrivial plasmoid-hierarchical substructure, is irrelevant for the particle in question. When the particle moves in such a layer, it only sees, and interacts with, EM fields that exist on length scales comparable to or larger than its Larmor radius in the reconnected field ($B_y$), while remaining essentially blind to the layer's smaller-scale substructure. Any EM structures on scales significantly smaller than the Larmor radius are smoothed out and are effectively invisible to the particle. 


In particular, as  a high-energy particle traverses the intermediate-scale inter-plasmoid layer (itself a plasmoid-mediated chain), it may encounter  many small secondary plasmoids with closed magnetic lines. However, the motion of the particle will not be strongly affected by its encounters with these plasmoids. Since the reconnected magnetic field in such a plasmoid reverses in the $x$-direction on a relatively small scale, it just deflects the particle a little bit in the $xz$ plane first one way and then the opposite way, with the two contributions cancelling each other. The particle will thus effectively pass right through the plasmoid. 
Likewise, even though the small plasmoids may have rapid motions that cause the $E_z$ field to fluctuate wildly, making it nonstationary and nonuniform as discussed in \S~\ref{subsec-picture-chain}, these small-scale electric fluctuations will not affect the particle's acceleration. As the particle passes through a small moving plasmoid, it will experience the $E_z$ field of one sign and then of the other, so the effects of the two halves of the plasmoid cancel each other, resulting in a zero net change.
To sum up, as long as a particle is unmagnetized on the scale of a small plasmoid it encounters in the course of its motion through a reconnecting plasmoid chain, such an encounter does not result in a significant change in the particle's regular and more or less steady acceleration by~$E_{\rm  rec}$. 

This process of regular acceleration by $E_{\rm  rec}$ proceeds for a while, but then there is a certain probability per unit time, or unit path length, for the particle to encounter a patch of reconnected field~$B_y$ that is strong and extended enough to magnetize it, i.e., 
\beq
\rho(\gamma, B_y) = {{\gamma m_e c^2}\over{eB}} =\gamma \rho_0\, {B_0\over B_y} \lesssim \Delta x \, , 
\eeq
where $\Delta x$ is the extent of the patch in the direction perpendicular to the reconnected field~${\bf B}_y$. 
A more general and precise (but basically equivalent) formulation of this condition is 
\beq
\epsilon \equiv \gamma m_e c^2 < e \Delta \Psi \, , 
\label{eq-magnetization_cond-flux}
\eeq
where $\Delta \Psi \equiv \Delta x  B_y$ is the magnetic flux (i.e., the $z$-component of the magnetic vector potential) of the patch. Interestingly, this condition is similar to that for electrostatic trapping in an electrostatic potential well of depth~$\Delta \varphi$, with the electrostatic potential $\Delta \varphi$  replaced by the drop in the vector potential $|\Delta A_z| = \Delta \Psi$. That is, particle trapping in plasmoids can be viewed as a magnetic analog of particle trapping in electrostatic potential wells of nonlinear electrostatic waves (electron holes, etc.).

When this happens, the particle may with some probability be reflected back into the acceleration region and continue accelerating, or it may be taken out of the active acceleration process, at least temporarily. 
In a plasmoid-dominated reconnection layer, the reconnected field $B_y$ can be thought of as having a more-or-less bimodal distribution: 

\smallskip
\noindent{\bf (i)} the relatively weak, distributed reconnected field of typical strength $B_1 \sim \epsilon B_0 \sim 0.1 B_0$ present almost everywhere in inter-plasmoid current layers outside of circularized plasmoids (see \S~\ref{subsec-picture-chain});  

\smallskip
\noindent {\bf (ii)} the $B_{\rm pl} \sim B_0$  field inside fully formed, circularized plasmoids with aspect ratios of order~1 \citep{Sironi_etal-2016} (this field can be even stronger deep inside plasmoid cores). 

\medskip
Because of this bimodal nature of the $B_y$-distribution, in our analysis we will treat the magnetization by $B_1$ (see \S~\ref{subsec-magnetization}) and particle interaction with large plasmoids as separate processes, with the latter further subdivided into particle trapping inside plasmoids (\S~\ref{subsec-trapping}) and particle reflection by moving plasmoids (\S~\ref{subsec-Fermi}). 
Here we present a qualitative physical discussion of these processes.

First, if a particle becomes magnetized by the $B_1$ field in an inter-plasmoid reconnection layer, it starts performing electric drift associated with $E_{\rm rec}$ and~$B_1$, thus moving with the general bulk plasma outflow in the $\pm x$-direction, away from that layer's X-point. Its acceleration then slows down dramatically.  However, since the reconnected magnetic field $B_y$ generally strengthens in the outflow direction, the particle's energy still increases gradually, responding adiabatically to the magnetic field compression while obeying the conservation of the particle's magnetic moment (the first adiabatic invariant), and also to the field-line shortening while preserving the second adiabatic invariant. Eventually, this stage ends when the particle reaches a big plasmoid at the end of the inter-plasmoid layer under consideration (see below); the particle then is either trapped by the plasmoid or kicked out back into the active acceleration zone where it can resume rapid acceleration. 
One may, therefore, examine whether it is possible for a particle that got caught in a patch of cross-layer reconnected magnetic field extended enough to magnetize it and thus inhibit its acceleration, to ever be ``released back into the wild" again. 
To investigate this question, let us suppose that the patch initially has a field~$\sim B_1$, representing the typical field at the edge of a reconnection-layer outflow, and is subsequently pulled and absorbed into an adjacent big magnetic island (plasmoid). This phase corresponds to the contraction and circularization of this newly-added part of the island, with the field strength in the patch increasing from $B_1$ to~$B_0$ while preserving its flux~$\Delta\Psi$.  Will the particle continue to be magnetized within this patch of reconnected flux? 
If this were indeed so, then during this contraction phase the particle's perpendicular energy would rise in betatron acceleration due to the conservation of the relativistic particle's magnetic moment, $\mu \sim \epsilon^2 /B = {\rm const}$. Thus, the particle's energy would increase by a factor of $(B_0/B_1)^{1/2} \sim 3$. Therefore, if the magnetization condition~(\ref{eq-magnetization_cond-flux}) was initially satisfied only marginally, i.e., $\epsilon \simeq e \Delta \Psi$, then it may no longer be satisfied by the end of this process; the particle may then escape from the patch under consideration, perhaps receiving a modest (order unity) energy boost. It may then again be able to enter another acceleration region, e.g, another inter-plasmoid reconnection layer, and be accelerated by $E_{\rm rec}$ once again. 
If, however, the plasmoid is sufficiently large then even if the particle is no longer magnetized to the patch in question, it would still continue to be trapped by this plasmoid. 

Next, let us consider what happens when an energetic particle encounters a fully formed, circularized plasmoid large enough to confine it.  This can happen either when the particle comes out unmagnetized from the inter-plasmoid current sheet and suddenly hits upon the plasmoid in question, or when it first gets magnetized by $B_1$ and then ${\bf [E\times B]}$-drifts along with the general outflow from the inter-plasmoid reconnection layer and gets pulled into the plasmoid by the contracting reconnected magnetic flux, as described above. 
In either case, the condition for the plasmoid to be large enough to be able to interact strongly (e.g., trap) with the particle is that the plasmoid width (i.e., radius, for a circularized plasmoid) $w$ is equal or larger than the particle's Larmor radius in the plasmoid's magnetic field~$B_{\rm pl}$: 
$w \geq w_{\rm tr}(\gamma) \equiv \rho_L(\gamma, B_{\rm pl})$.
Assuming a circularized plasmoid of reasonable size, we can simply take $B_{\rm pl} \simeq B_0$, and thus write this condition as
$w \geq w_{\rm tr}(\gamma) \equiv \rho_L(\gamma, B_0) = \gamma\rho_0 = \gamma m_e c^2/e B_0$. 
According to~(\ref{eq-magnetization_cond-flux}), this condition can also be recast equivalently in terms of the plasmoid's flux $\Psi \simeq B_0 w$ as $\epsilon = \gamma m_e c^2 \lesssim e \Psi$. 
We will call any plasmoid that satisfies this condition, a {\it large plasmoid}. 
Thus, in our terminology, ``{\it large plasmoid}" is a technical term, defined not in some vague absolute sense but only in relation to a particle of a given energy.  For example,  a typical, average-energy ($\gamma\sim \bar{\gamma}$) particle attains its energy while being accelerated over only one elementary layer at the bottom of the plasmoid hierarchy.  Then, any plasmoid of width~$w$ larger than the elementary layer thickness~$\delta$, which in collisionless reconnection without a strong guide field is comparable to the average particle Larmor radius , $\delta \sim \bar{\rho}$, is considered to be a large plasmoid for such a particle. If, however, we consider an energetic particle in the nonthermal tail, with energy, $\gamma \gg \bar{\gamma}$, then the smallest ``large" plasmoid that can trap this particle is somewhere in the middle of the plasmoid hierarchy.  Consequently, the intermediate reconnection layer between two neighboring such large plasmoids is, in general, not an elementary layer but itself a hierarchical plasmoid chain, with a number of smaller plasmoids and reconnecting X-points between the two large plasmoids. Thus, an energetic particle may cross multiple small plasmoids and inter-plasmoid current layers while on its way to becoming eventually trapped in a large plasmoid.

When a particle finally encounters a large plasmoid, one of three things can happen.  
First, the particle may just go around it (by circling it in the $xy$-plane) and continue its motion essentially unchanged in the same direction on the other side, with no substantial energy change. 
 
Second, there is a finite probability that the particle will get captured and trapped inside the plasmoid. 
The particle is then essentially removed from the active acceleration zone, at least for some substantial interval of time. 
Thus, perhaps counter-intuitively, in our model particle {\it trapping} in plasmoids plays the role of a particle {\it escape} mechanism in the language of Fermi acceleration \citep[see, e.g.,][]{Werner_etal-2018}. 
In other words, a particle ``escapes" from the acceleration process not by flying away, leaving the reconnection region altogether, but by leaving the active acceleration zone within the reconnection region by being trapped and sequestered inside a quiet large plasmoid.   

As mentioned above, a particle with a large energy may pass many small plasmoids before it meets a plasmoid large enough to magnetize and trap it. Because such large plasmoids are rare, the particle will cover a rather large distance before meeting one; and since it is being continuously accelerated during all this time, its energy and hence its Larmor radius will grow substantially. The interplay between this continuous Larmor-radius growth and trapping in large plasmoids ties the spectrum of accelerated particles to the distribution function of plasmoids and ultimately determines the efficiency, and the high-energy cutoff of~NTPA. 

Once trapped, a particle effectively stops accelerating, at least for a while, and is just carried around inside a large plasmoid. More precisely, it no longer undergoes a continuous vigorous, rapid acceleration by $E_{\rm rec}$ or by Fermi acceleration.   
Nevertheless, it can still get a further increase in energy via a number of ways. 
First, it has a chance to be reenergized if its host plasmoid undergoes a merger with another plasmoid \citep{Oka_etal-2010, Sironi_Spitkovsky-2014, Nalewajko_etal-2015, Sironi_etal-2016, Li_etal-2017}. 
For example, there is numerical evidence that the most energetic particles accelerated by reconnection gain their energy in a two-stage process: first, pre-accelerating by the main reconnection electric field in an inter-plasmoid reconnection layer and then getting an additional energy boost in a plasmoid merger \citep{Sironi_Spitkovsky-2014}.
Since such a merger is also a reconnection event (sometimes called ``anti-reconnection" because the corresponding reconnection electric field points in the direction opposite to $E_{\rm rec}$ in the main reconnection layer), further particle acceleration is expected. 
The key parameters of this secondary reconnection event, however, differ from those of the original, primary reconnection. First, the length of the perpendicular secondary reconnection layer, roughly the size of the smallest of the two merging plasmoids, and is typically shorter than the length of the inter-plasmoid reconnection layer where the particle had been accelerated before it was captured by one of the plasmoids. 
This, in turn, implies that the reconnecting magnetic flux and hence the maximum electric potential drop involved in the secondary reconnection event are also smaller than those in the primary reconnection event (and certainly less than those corresponding to the global reconnection layer).
In addition, since roughly one half of the upstream magnetic energy density has been converted into the plasma energy density during the primary reconnection process that has led to the creation of the two merging plasmoids, the energy content of these plasmoids is split roughly equally between the magnetic field and the plasma internal energy. Therefore, this secondary reconnection event generically takes place at a plasma $\beta$ of order unity (and hence, for an ultrarelativistic plasma, a hot magnetization~$\sigma_h$ of order unity), which further limits the effectiveness of the secondary particle acceleration.%
\footnote{We also would like to note that there are interesting similarities, which should be explored further, between hierarchical plasmoid mergers and galaxy mergers in the process of hierarchical structure formation.}

Another way in which energetic particles trapped inside a large circularized plasmoid may be energized further, even without plasmoid mergers, is the additional gradual adiabatic heating caused by the host plasmoid's inner core's readjustment and contraction in response to the plasmoid's growth on the outside  \citep{Petropoulou_Sironi-2018}. Long after a particle gets trapped on some closed flux surface (a closed field line loop in 2D reconnection without guide field) inside a large plasmoid, the plasmoid continues to grow by accretion and minor mergers with smaller plasmoids, piling up fresh accumulated mass and reconnected magnetic flux. This leads to a continuous readjustment of the plasmoid's internal magnetic structure, and the given particle's flux surface gets buried deeper and deeper inside. 
The magnetic pinch force of the newly accreted outer flux slowly compresses the plasma and the magnetic field in the inner part (the core) of the plasmoid, so that the closed magnetic field line that the particle circles becomes shorter while the field strength increases.  As a result, the conservation of both 1st (magnetic moment) and 2nd adiabatic (bounce; in this case, the angular momentum of the particle going around along the circular closed field line) invariants leads to a gradual increase of the perpendicular (due to field strengthening) and parallel (due to line shortening) particle energy, respectively. This process is, however, relatively slow, with the particle's energy increasing perhaps as $t^{1/2}$ \citep{Petropoulou_Sironi-2018}; hence we shall it ignore it in our analysis. 

This energization mechanism can operate even when the plasmoid trapping the particle is not yet fully circularized --- or, more precisely, when the particular closed flux surface on which the particle is trapped is not yet fully circularized, i.e., if it is still elongated in the $x$-direction.  Then, as this flux surface contracts and becomes more circular (while preserving the mass-per-flux ratio enclosed by it), it shortens, while the $B_y$ magnetic field strength at the intersection of this flux surface with the reconnection layer midplane ($y=0$) increases.  As the now-magnetized energetic particle moves along this surface, it may or may not find itself mirror-trapped to an area near the $y=0$ midplane because the magnetic field is weakest there and increases along the surface, attaining a maximum at the largest~$|y|$ (approximately directly above or below the plasmoid's O-point).%
\footnote{In addition, there may be electrostatic trapping of electrons along the field lines in the case electron-ion plasma reconnection.}
If the particle is trapped along the field in this way, it will gain energy due to both 1st (magnetic moment) and, in some cases, 2nd (bounce-motion) adiabatic invariant conservation. And if instead the particle is passing, not trapped by the mirror force, and is able to fully circulate around the plasmoid, it will also gain energy as its host flux surface contracts, again due to the 2nd adiabatic invariant conservation, in a Fermi acceleration process associated with the converging motion of the plasmoid's edges \citep{Drake_etal-2006}.  The energy ultimately comes from the work done on the particle by the out-of-plane ($E_z$) motional (ideal) electric field associated with the contracting motion of the field lines; the particle's $z$-displacement, necessary for this electric field to be able to do work on the particle, can be attributed to the particle's curvature drift as it turns around the contracting plasmoid's edges in its parallel motion \citep{Drake_etal-2006,  Dahlin_etal-2014}. 
Strictly speaking, one can invoke the 2nd adiabatic invariant conservation only in the case of non-relativistic reconnection, $V_A \ll c$, because then the edges of the flux surface, contracting at most at roughly the Alfv\'en speed, move much slower than the relativistic particle, and hence the particle can circle around the plasmoid (or undergo many bounces if it is mirror-trapped) many times during this contraction process, justifying the adiabatic assumption. In contrast, in relativistic reconnection, $V_A \sim c$, the Alfv\'enic contraction of the flux surface is itself relativistic, and the large time-scale separation required for the adiabatic description is lost --- the plasmoid contracts on the same time scale as the particle moves around it. Nevertheless, even if the mathematical language of the 2nd adiabatic invariant conservation and the associated Fermi acceleration is not applicable, the particle still gains energy in this process. 

\begin{figure}
\begin{center}
\includegraphics[width=10 cm]{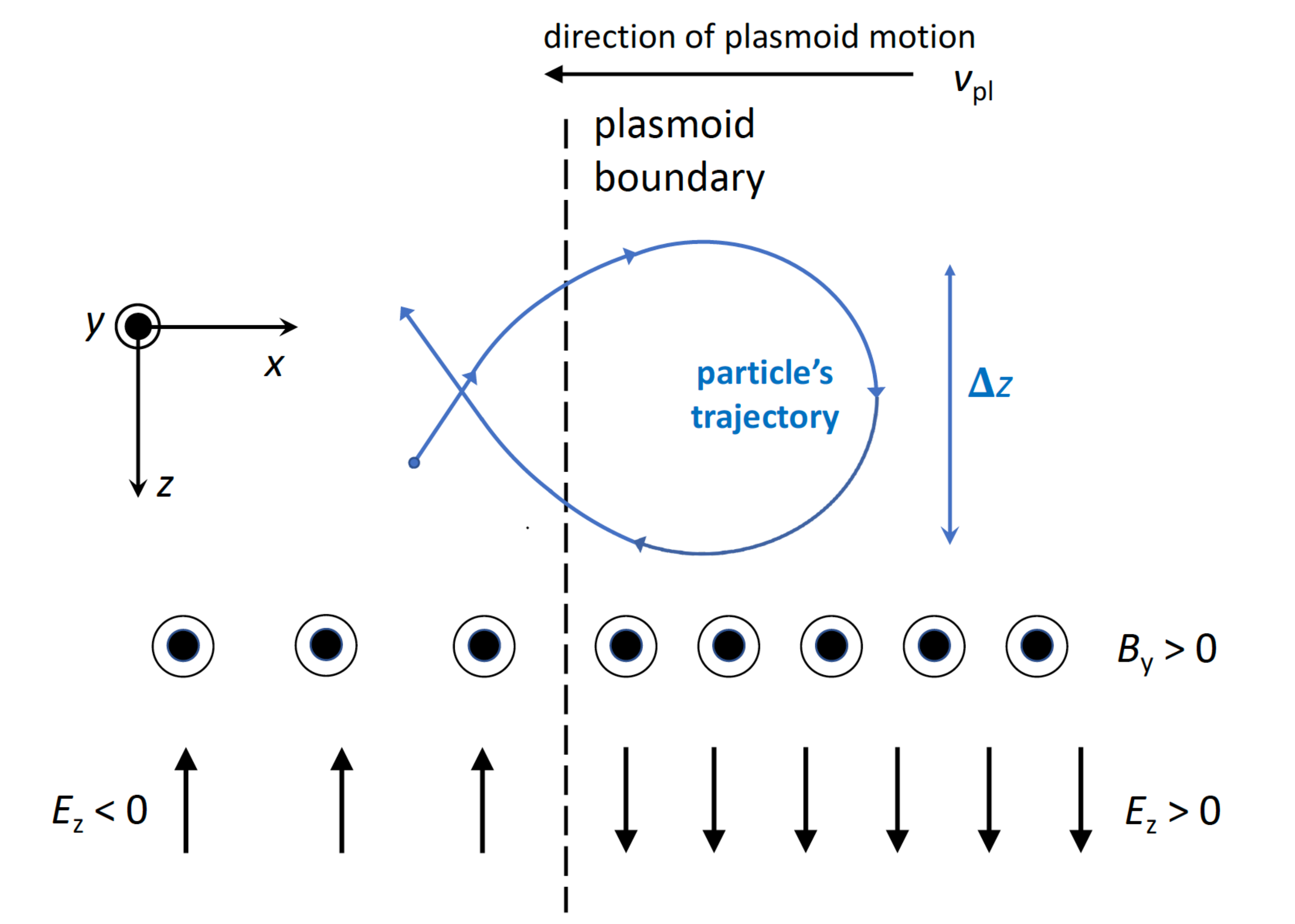}
\end{center}
\caption{Trajectory of an energetic positively-charged particle (shown in blue) initially moving to the right (positive-$x$) direction in the $xz$-plane (the current layer mid-plane) as it is being reflected by a large plasmoid moving to the left (negative-$x$ direction).  The vertical dashed line shows the plasmoid's left boundary. As the particle's $z$-component of motion is reversed by the strong reconnected magnetic field $B_y\sim B_0$ (out of the page) inside the plasmoid (to the right of the vertical dashed line), is maintains its alignment with the plasmoid's motional electric field, and thus continuously gains energy.}
\label{fig-3}
\end{figure}

The third possible outcome of an energetic particle's encounter with a large plasmoid is that  it can bounce off the plasmoid and get back into the main acceleration region; during this bounce, however, the particle interacts with the motional electric field associated with the plasmoid's motion as a whole. 
This interaction is brief, of order the gyro-period of the particle in the $B_y$ magnetic field of the plasmoid, which for circularized plasmoids is of order $B_0$; however, this interaction can be quite intense because the electric field here is~$\sim V_A B_0/c$. 
As a result, the particle may gain or lose a substantial amount of energy. 
Namely, assuming that an individual reflection is elastic in a moving plasmoid's frame, when considered in the lab frame, the particle can either gain or lose energy to the plasmoid depending on whether the plasmoid moves (in the $x$-direction) head-on or tail-on relative to the particle. 
From the microscopic point of view, this interaction can be understood as follows (see figure~\ref{fig-3}). 
Let us for definiteness consider a positively charged particle approaching a large plasmoid from the left, i.e., emerging from the inter-plasmoid layer located to the left of the plasmoid and moving to the right (positive $x$, say), i.e., in the direction of the general reconnection outflow.  The plasmoid, however, can be moving either to the left, opposite to the particle's direction of motion (a head-on collision), or to the right, in the same direction as the particle (a tail-on collision). 
If the interaction is head-on (as in figure~\ref{fig-3}), the plasmoid moves in the direction opposite to the overall large-scale reconnection outflow, and, because $B_y$ in the left half of the plasmoid (facing the approaching particle) is in the same direction as in the adjacent inter-plasmoid current sheet, the motional electric field ($E_z$) in this part of the plasmoid is opposite to the main reconnection electric field~$E_{\rm rec}$. Then, when the particle approaching from the left encounters this plasmoid, it gets partially magnetized by it for a short period of time: the plasmoid's $B_y$ field deflects the particle so that it performs an incomplete gyro-orbit (more than a half but less than full) inside this plasmoid, thereby reversing its $x$-motion, and then escapes from the plasmoid back into the inter-plasmoid current layer from which it came. 
Effectively, the particle is reflected by the plasmoid's magnetic field. 
Importantly, while the particle is covering this incomplete gyro-orbit inside the plasmoid, it is moving backwards in $z$ relative to its $z$-motion prior to the encounter. Thus, both the particle's $z$ direction of motion and the electric field are reversed relative to what they were in the main current-layer acceleration zone; therefore, they are again aligned and so that the particle gains energy from the plasmoid (see \S~\ref{subsec-Fermi} for quantitative details). This process is a special version of gyro-resonance acceleration over one gyro-orbit: the electric field and the particle's velocity ($z$ components) stay aligned during one gyro-orbit. 


In the opposite case (a tail-on collision), when the plasmoid in moving to the right, the reconnected magnetic field on its left side (interacting with the particle) is still in the same direction as in a head-on collision, and so the particle gets temporarily magnetized for $\gtrsim$ half of a gyro-orbit and reflected by the plasmoid leftward, back into the layer, in a way similar to a head-on collision. However, the direction of the motional $E_z$ electric field in this part of the plasmoid is now the same as, not opposite to, that in the inter-plasmoid layer on the left. Hence, this electric field and the backward $z$-motion of the particle in its half gyro-orbit inside the plasmoid are now counter-aligned and thus the particle loses energy in the encounter.  

The net effective result of many such encounters is \cite{Fermi-1949} acceleration. 
If plasmoid motions are random, uncorrelated, then one gets second-order, diffusive Fermi acceleration (considered in \S~\ref{subsec-Fermi}). However, if a particle bounces repeatedly between two approaching large plasmoids, the energy kicks are strongly correlated and the particle rapidly gains energy in a first-order Fermi acceleration.

In summary, we would like to reiterate that here we try to avoid concepts like parallel or perpendicular electric field acceleration. 
The particles are primarily accelerated by the out-of-plane reconnection electric field, ${\bf E}_z$. In the absence of a guide magnetic field, as is the main focus of the paper, and of the quadrupolar out-of-plane Hall magnetic field that arises in collisionless electron-ion plasma reconnection, the accelerating electric field is naturally perpendicular to the magnetic field almost everywhere. 
But this fact may not be very important; the exact orientation of the electric field relative to the actual local, microscopic magnetic field (e.g., whether it is parallel or perpendicular) is not particularly relevant to high-energy particles. A given energetic particle is not sensitive to EM fields  with scales much smaller than its Larmor radius. 
In particular, even though the electric field is perpendicular to the local magnetic field, this does not necessarily mean that its acceleration is due to a drift motion (such as, e.g., the curvature drift) along the perpendicular electric field, since, in order to describe a particle's motion as a drift, the particle needs to be magnetized in the first place. 


\section{Theoretical Model}
\label{sec-model}


In our picture we split the whole particle population into two sub-populations: (1) particles currently undergoing active acceleration and (2) particles that are captured by the reconnected magnetic field and eventually trapped inside large plasmoids.  The overall flow of particles is from the upstream region into the actively reconnecting current layers, where they can experience significant acceleration, and from there into their comfortable, quiet ``retirement" inside plasmoids (although some particles go directly from the upstream into the plasmoids).  In 2D it is relatively rare for a particle already trapped deep in a large plasmoid to escape back into an active acceleration zone; this only happens occasionally during plasmoid mergers.  The discussion in this paper will focus on the actively accelerating particles outside plasmoids. Since these particles are ultimately injected into plasmoids, the resulting particle energy distribution function~$f(\gamma)$ studied in this paper can provide the source term for theories concerned with the evolution of the particle population trapped inside plasmoids, such as that presented in the recent paper by~\cite{Hakobyan_etal-2020}. 


\subsection{Kinetic Equation for NTPA in a Plasmoid-Dominated Reconnection Layer}
\label{subsec-kinetic-eqn}

We describe relativistic particle acceleration by reconnection quantitatively in terms of a kinetic equation governing the energy distribution function~$f(\gamma)$ of the particles in the acceleration zone: 
\beq
\partial_t f(\gamma,t) = S(\gamma) 
- \, \partial_\gamma (\dot{\gamma}_{\rm acc} f) - {f(\gamma)\over{\tau(\gamma)}} + 
\partial_\gamma (D_\gamma \partial_\gamma f)\, .
\label{eq-kinetic-1}
\eeq
Here, the first term on the  right-hand side (RHS), $S(\gamma)$, is the source term describing the injection of the particles into the reconnection layer from the upstream region; $\dot{\gamma}_{\rm acc}$ in the next term is the rate of regular, continuous acceleration by the reconnection electric field~$E_{\rm rec}$; the third term, $-f(\gamma)/\tau$, represents the effective escape of particles from the acceleration process by their capture onto cyclotron orbits around the reconnected field~$B_y$; and the last term represents diffusive Fermi acceleration due to the particle bouncing off moving plasmoids. 

We shall now make several simplifications. 
First, we will focus on the quasi-steady-state solution, which is appropriate for the subpopulation of energetic particles undergoing active, rapid acceleration before they are taken out of this process by being magnetized by the reconnected magnetic field and trapped inside plasmoids.  In contrast, the subpopulation of energetic particles trapped in plasmoids grows continuously and is thus, of course, not stationary; it is being constantly fed by the escape (the $f/\tau$ term) of particles from the actively accelerating subpopulation under consideration here; however, we relegate the analysis of this trapped subpopulation to future studies (see, e.g., \citealt{Hakobyan_etal-2020}).  
As for the actively accelerating particles, it is reasonable to expect that they may develop a quasi-stationary spectrum governed by the balance between the injection $S(\gamma)$ at small energies, a flow up in energy due to the regular reconnection acceleration $\dot{\gamma}_{\rm acc}$ and Fermi acceleration by moving plasmoids [the second and last terms in~(\ref{eq-kinetic-1})], and escape, $-f/\tau$.

Our second simplification is based on the assumption that the background upstream plasma is cold (magnetically-dominated, $\beta_b \ll 1$). Then, the source term is concentrated at small energies characteristic of the upstream conditions, $\gamma_{\rm inj}$, much less than the energies $\bar{\gamma} \sim \sigma_c$ of accelerated particles of interest to us here.  We will therefore ignore the injection term in our analysis of acceleration of high-energy nonthermal particles; this is similar in spirit to ignoring the details of energy injection at the large driving scale when analyzing the inertial range of turbulence. 

The steady-state kinetic equation (\ref{eq-kinetic-1}) at $\gamma \gg \gamma_{\rm inj}$ then becomes 
\beq
- \, \partial_\gamma (\dot{\gamma}_{\rm acc} f) - {f(\gamma)\over{\tau(\gamma)}} + \partial_\gamma (D_\gamma \partial_\gamma f)\, = 0  \, .
\label{eq-kinetic-2}
\eeq

Next, as discussed in \S~\ref{sec-picture}, the reconnected magnetic field $B_y$ has a roughly bimodal distribution in a reconnecting plasmoid chain, corresponding to two types of regions: inter-plasmoid current layers with a relatively weak distributed reconnected field of order $B_1 \sim \epsilon B_0$, and circularized plasmoids where the field is of order~$B_0$. 
To reflect this dichotomy, it is convenient to split the escape term in the kinetic equation into two separate terms encapsulating the two effective escape channels that check the continuous particle acceleration: magnetization in the general reconnected field $B_1$ and trapping in large plasmoids (see \S~\ref{sec-picture}): 
\beq
{1\over \tau(\gamma)} = {1\over \tau_{\rm magn}(\gamma)} + {1\over \tau_{\rm trap}(\gamma)} \, .
\label{eq-tau}
\eeq


We will now discuss the individual terms in equation (\ref{eq-kinetic-2}) one by one to better understand their underlying physics and the role they play in shaping up the particle distribution. 
For most of this discussion, however, for the sake of simplicity and analytical tractability, we will ignore the diffusive acceleration term. 
We will come back to it only in \S~\ref{subsec-Fermi}, where we will evaluate its importance in different regimes and will obtain an explicit analytical solution in one special case. 
We will then discuss the general kinetic equation in~\S~\ref{subsec-general_kin_eq}.


\subsection{Regular acceleration by the reconnection electric field} 
\label{subsec-accel}

Consider a particle that is moving in the layer before it gets magnetized by the reconnected field~$B_1$ or trapped in a big plasmoid.  The particle then undergoes a regular, steady acceleration by the main reconnection electric field $E_{\rm rec} = \epsilon \beta_A B_0$, with the acceleration rate given by 
\beq
\dot{\gamma}_{\rm acc} = \tau_{\rm reg}^{-1} = 
{{e E_{\rm rec} v_z}\over{m_e c^2}} \sim \beta_z \Omega_0 {E_{\rm rec}\over{B_0}} = \beta \Omega_0 \epsilon  {V_A\over{c}}
=  \epsilon \beta_z \beta_A \Omega_0 \, ,
\eeq
where $\Omega_0 \equiv e B_0/m_e c$ is the nominal non-relativistic electron cyclotron frequency and $\beta_z \equiv v_z/c$.
As long as the particle is not yet strongly deflected in the $x$ direction by the Lorentz force due to the reconnected field $B_y$ (which is related to the condition that it is not magnetized by this field), it should have a finite, sizable velocity component in the $z$ direction, so that $\beta_z \simeq 1$ (assuming the particle is ultra-relativistic). 
Importantly, we then see that the acceleration rate (\ref{eq-gamma-dot}) is independent of the particle energy, 
\beq
\dot{\gamma}_{\rm acc} \simeq  \epsilon \beta_A \Omega_0 = {\rm const}.
\label{eq-gamma-dot}
\eeq

Substituting this expression into the kinetic equation~(\ref{eq-kinetic-1}) without the diffusion term, we see that a steady-state distribution is governed by
\beq
\dot\gamma_{\rm acc} \, {{{\rm d} f}\over{{\rm d} \gamma}} = 
- \, {{f(\gamma)}\over{\tau(\gamma)}} \quad \Rightarrow \quad 
{{{\rm d} \ln f}\over{{\rm d} \gamma}} =  - \, {1\over{\dot\gamma_{\rm acc} \tau(\gamma)}} \simeq
- \, {1\over{\epsilon \beta_A \Omega_0 \tau(\gamma)}} \, . 
\label{eq-kin-steady}
\eeq

Integrating equation (\ref{eq-kin-steady}), we obtain the stationary distribution function as
\beq
f(\gamma) = 
C\, \exp \biggl(-{1\over{\dot\gamma_{\rm acc}}}\, 
\int\limits^{\gamma} {{\rm d}\gamma'\over{\tau(\gamma')}}\biggr) \simeq  
C\, \exp \biggl(-{1\over{\epsilon \beta_A \Omega_0}}\, 
\int\limits^{\gamma} {{\rm d}\gamma'\over{\tau(\gamma')}}\biggr)\, .
\label{eq-soln-steady-general}
\eeq

Recalling (\ref{eq-tau}), this distribution function can be represented as a product of two factors, one due to magnetization and one due to trapping: 
\beq
f(\gamma) = 
C\, \exp \biggl(-{1\over{\dot\gamma_{\rm acc}}}\, 
\int\limits^{\gamma} {{\rm d}\gamma'\over{\tau_{\rm magn}(\gamma')}}\biggr) 
\times \exp \biggl(-{1\over{\dot\gamma_{\rm acc}}}\, 
\int\limits^{\gamma} {{\rm d}\gamma'\over{\tau_{\rm trap}(\gamma')}}\biggr) \, .
\label{eq-soln-steady-2-factors}
\eeq


We will now proceed to discuss the two escape terms on the RHS of equation (\ref{eq-kinetic-2}). As we will see, these two terms affect the mathematical shape of the resulting distribution function in different ways, because their characteristic time- and  length-scales have different scalings with the particle energy~$\gamma m_e c^2$. Thus, $\tau_{\rm magn}$ is basically directly proportional to~$\gamma$, while $\tau_{\rm trap}$ is tied to the plasmoid distribution function. As we will show below, the first term controls the power-law slope and the second one controls the high-energy cutoff.


\subsection{Particle magnetization by inter-plasmoid reconnected magnetic field $B_1$: governing the power-law index} 
\label{subsec-magnetization}

As a particle moves through a reconnection region and experiences acceleration by the $E_{\rm rec}$ field, it also interacts with the reconnected magnetic field $B_y \sim B_1 \simeq \epsilon B_0$, which continuously deflects it more and more towards $\pm x$ direction, out of the accelerating layer. 
At some point, after the particle travels a certain distance $l_{\rm magn}$ in the $x$-direction,  this deflection may become so large that the particle becomes effectively magnetized by this field. Its subsequent motion in the $xz$ plane then becomes dominated by the cyclotron motion associated with the field~$B_1$, coupled with ${\bf E \times B}$ drift in the $x$-direction, which corresponds to the particle moving with the general (fluid-level) reconnection outflow. The resulting inability of the particle to move unimpeded along the reconnection electric field (in the $z$ direction) means that its rapid energy gain becomes greatly diminished and its acceleration slows down.%
\footnote{As discussed in \S~\ref{subsec-picture-particle-acceleration}, however, the magnetized particle can still gain energy slowly, e.g., by betatron acceleration as it drifts into a region of stronger~$B_y$.}  
This phase continues until the particle encounters a large plasmoid, at which point it will have a finite chance to get absorbed into it (see \S~\ref{subsec-trapping} below).  

The distance $\ell_{\rm magn}(\gamma)$ that a particle of a given energy $\gamma m_e c^2$ has to travel in the $x$-direction before becoming effectively magnetized can be estimated simply as the Larmor radius corresponding to~$B_1$ \cite[e.g.][]{Zenitani_Hoshino-2001}:  
$\ell_{\rm magn} (\gamma) \sim \rho_L(\gamma, B_1) \sim (B_0/B_1) \rho_L(\gamma, B_0) \sim \epsilon^{-1} \rho_L(\gamma, B_0)  = \epsilon^{-1} \gamma \rho_0 \sim 10 \gamma \rho_0$.  
Since a particle in the acceleration region typically moves with a finite angle with respect to the $z$ axis, so that $v_x \sim v_z \sim c$, we can estimate the time for a particle to get magnetized as
\beq
\tau_{\rm magn}(\gamma) \sim \ell_{\rm magn} /c  \sim \epsilon^{-1} \gamma \Omega_0^{-1} \, . 
\label{eq-tau_magn}
\eeq

Furthermore, the typical amount of energy that the particle gains by regular acceleration while it crosses the distance $\ell_{\rm magn}$ before becoming magnetized is of order
$\delta \gamma (\ell_{\rm magn}) \sim e E_{\rm rec} \ell_{\rm magn}/m_e c^2 = (E_{\rm rec}/B_0) \ell_{\rm magn}/\rho_0 \sim \gamma (E_{\rm rec}/\epsilon B_0)  \sim \gamma (E_{\rm rec}/B_1) \sim \gamma \beta_A$. 
Thus, the fractional energy gain, $\delta \gamma (\ell_{\rm mag})/\gamma \sim \beta_A$, is of order unity in the case of relativistic reconnection [$\sigma_h > 1$ and hence $\beta_A\simeq 1$, see~(\ref{eq-V_A})], but becomes small, of order $\beta_A \sim \sigma_h^{1/2}$ in the non-relativistic reconnection case.  

%

Importantly, the magnetization time (\ref{eq-tau_magn}) is directly proportional to the particle's energy, and this  enables a power-law distribution to develop.  
Indeed the corresponding factor in equation (\ref{eq-soln-steady-2-factors}) is 
\beq
\exp \biggl(-\, {1\over{\dot\gamma_{\rm acc}}}\, \int {{\rm d}\gamma\over{\tau_{\rm magn}(\gamma)}}\biggr)  = 
\exp \biggl(-\, {1\over{\epsilon \beta_A \Omega_0}} \, \int {\epsilon \Omega_0 {\rm d}\gamma \over{\gamma}} \biggr) = 
\exp \biggl(-\, {1\over{\beta_A}} \, \int\limits^{\gamma} {{\rm d}\gamma' \over{\gamma'}} \biggr) \sim 
 \gamma^{-p} \,  , 
\eeq
where the power-law index is given by 
\beq
p = p_{\rm magn}(\sigma_h) \sim {1\over \beta_A} = \sqrt{{1+\sigma_h}\over{\sigma_h}}  \, .
\label{eq-p-magn}
\eeq

Thus we see that the balance between regular acceleration by the reconnection electric field $E_{\rm rec} \sim \epsilon \beta_A B_0$ and particle magnetization by the reconnected magnetic field $B_1\sim \epsilon B_0$ produces NTPA with a power-law index $p_{\rm magn}$ that exhibits the same dependence on the hot magnetization~$\sigma_h$ as was observed in recent numerical PIC studies \cite[e.g.,][]{Sironi_Spitkovsky-2014, Guo_etal-2014, Guo_etal-2015, Werner_etal-2016, Werner_etal-2018, Ball_etal-2018}, namely: 
\smallskip
\begin{itemize}
\item $p \rightarrow {\rm const}$  of order unity in the ultra-relativistic reconnection limit, 
$\sigma_h \gg 1$, $V_A \simeq c$; 
\item $p \sim \sigma_h^{-1/2}$ in the nonrelativistic case, $\sigma_h \ll 1$ and hence $V_A \simeq c\sigma_h^{-1/2} \ll c$.
\end{itemize}

\smallskip
We also note that this physical picture and the theoretical arguments are similar to those presented by \cite{Zenitani_Hoshino-2001} for a simple laminar (without secondary plasmoids) reconnection layer in the ultra-relativistic limit.


\subsection{Particle Trapping in Plasmoids and the High-Energy Cutoff: General Discussion}
\label{subsec-trapping}

We shall now discuss the effects of the second factor in equation (\ref{eq-soln-steady-2-factors}) --- particle trapping in large plasmoids --- and will argue that in large systems, where reconnection proceeds in the plasmoid-mediated regime, this process controls the extent of the power-law segment of the particle energy distribution; in particular, it induces an exponential-like high-energy cutoff~$\gamma_{c}$. 
Thus,  in our model the particle spectrum at highest energies is shaped by the distribution of plasmoids.
We will postpone the discussion of the other important aspect of particle-plasmoid interaction, i.e., the Fermi acceleration by particle reflections off rapidly moving plasmoids, until \S~\ref{subsec-Fermi}.

As discussed in \S~\ref{subsec-picture-particle-acceleration}, we use the term {\it ``large plasmoid"} as a technical term with a specific meaning: a plasmoid that is large enough to trap a particle of a given energy~$\gamma m_e c^2$, i.e., a plasmoid with the size  comparable to, or larger than, the particle's Larmor radius in the plasmoid's magnetic field.  
Also as discussed in \S~\ref{subsec-picture-particle-acceleration}, the trapping condition is actually most accurately cast in terms of the plasmoid's magnetic flux~$\psi$, i.e., the absolute value of the difference in the out-of-plane ($z$) component of the electromagnetic vector potential $A_z$ between the plasmoid's center (i.e., the O-point) and its edge: $\epsilon  = \gamma m_e c^2 < e \psi$.
Thus, strictly speaking, we should be dealing with the plasmoid distribution function with respect to their fluxes.  
However, while the formulation of the theory in terms of plasmoid fluxes is more rigorous, its formulation in terms of sizes  $w$ is arguably more intuitive and easier to visualize, and so this is the language we will use in this paper. 

Once fully formed, plasmoids tend to be roughly circularized, with aspect ratios of order~1, and with characteristic magnetic fields comparable to~$B_0$.%
\footnote{For simplicity, here we shall ignore the internal structure of plasmoids and, in particular, the fact that the magnetic field can be substantially compressed inside them, especially in the no-guide-field case and for relativistically hot plasmas, which are more compressible than non-relativistic plasmas due to their lower adiabatic index (4/3 instead of~5/3). We note, however, that in the version of our theory cast in terms of plasmoid fluxes instead of sizes this issue does not arise.}
Then, as discussed in \S~\ref{subsec-picture-particle-acceleration}, we postulate a 1-to-1 correspondence between plasmoid fluxes and sizes, $\psi = w B_0$, and then the trapping condition can be written as
\beq
w \geq w_{\rm tr} (\gamma) \equiv \rho_L(\gamma, B_0)  = \gamma \rho_0  \, .
\eeq

Correspondingly, there exists a characteristic distance that a given particle is able to travel in the $x$-direction before it is trapped by a large plasmoid: this is the characteristic separation $\lambda_{\rm pl}[w_{\rm tr}(\gamma)]$ between plasmoids of this size~$w_{\rm tr}(\gamma)$. 
This separation is, in turn is controlled by the plasmoid-size distribution function in the reconnecting plasmoid chain, i.e.,
\beq
\lambda_{\rm pl}(w)  = {L\over{N(w)}} \, .
\label{eq-lambda-N}
\eeq
Here $L$ is the global length of the layer, and $N(w)$ is the number of plasmoids with size equal or greater than~$w$, i.e., the cumulative plasmoid-size distribution function. It is related to the plasmoid distribution density~$F(w)$ as 
\beq
F(w) = - dN/dw \, , 
\eeq
i.e., 
\beq
N(w) = \int\limits_w F(w') dw'  + {\rm const} \, .
\eeq

Using equation (\ref{eq-lambda-N}), the characteristic trapping time $\tau_{\rm trap}(\gamma)$ that enters the kinetic equation~(\ref{eq-kinetic-2}) can be estimated as 
\beq
\tau_{\rm trap}(\gamma) \sim {\lambda_{\rm pl}[w_{\rm tr}(\gamma)]\over{c}} 
= {L\over c} \, {1\over N[w_{\rm tr}(\gamma)]}\, , 
\label{eq-tau_trap}
\eeq
where we have assumed that a relativistic particle's motion through the acceleration region before it is trapped or magnetized is generally at a finite angle with respect to the $z$-axis, so that its typical velocity in the $x$-direction (i.e., along the layer) is relativistic, $v_x \sim c$. 
From~(\ref{eq-tau_trap}) we see that higher-energy particles take longer to get trapped because they need to encounter a large enough plasmoid, which is rare. 

The corresponding term in the kinetic equation is
\beq
- \, {f(\gamma)\over{\tau_{\rm trap}(\gamma)}} \sim -\, f \,{c\over{\lambda_{\rm pl}[w_{\rm tr}(\gamma)]}}  \sim 
-\, f {c\over{L}} N[w_{\rm tr}(\gamma)] \, ,
\eeq
and the resulting suppression factor in the particle energy distribution function (\ref{eq-soln-steady-2-factors}) is
\begin{eqnarray}
 \exp \biggl(-\, {1\over{\dot\gamma_{\rm acc}}}\, 
 \int\limits^{\gamma} {{\rm d}\gamma'\over{\tau_{\rm trap}(\gamma')}}\biggr) 
 &=&  \exp \biggl(-\, {1\over{\epsilon \beta_A \Omega_0}}\, {c\over L} 
 \int\limits^{\gamma}  N[w_{\rm tr}(\gamma')] {\rm d}\gamma' \biggr) \nonumber \\
 &= & \exp \biggl(-\, {1\over{\epsilon \beta_A}}\, {\rho_0\over L} 
 \int\limits^{\gamma} N[w_{\rm tr}(\gamma')] {\rm d}\gamma' \biggr)  \, .
\label{eq-trap}
\end{eqnarray}
These expressions explicitly embody the relationship between particle acceleration and plasmoid distribution~$N(w)$. 

%

In order to evaluate the effect of this factor on particle acceleration in more concrete terms, we need a model for the plasmoid-size distribution function. 
For illustration, let us first consider a chain of equidistant identical plasmoids of a single size $w_*$ separated by a distance $\lambda_*$; thus, the number of such plasmoids in the chain is $N_* = L/\lambda_*$.  The plasmoid distribution density then is $F(w) = N_* \delta(w-w_*)$ and the cumulative distribution function is given by the Heaviside step-function: 
$N(w) = N_* [1- \Theta(w-w_*)]$. 
These plasmoids can trap all particles with energies $\gamma \leq \gamma_* \equiv w_*/\rho_0$, and cannot trap particles with higher energies. 
Then the resulting suppression factor for particles with $\gamma \leq \gamma_*$ is 
\begin{eqnarray}
 \exp \biggl(-\, {1\over{\epsilon \beta_A}}\, {\rho_0\over L} 
 \int\limits^{\gamma} N[w_{\rm tr}(\gamma')] {\rm d}\gamma' \biggr)  
 &=&  \exp \biggl(-\, {1\over{\epsilon \beta_A}}\, {N_* \rho_0\over L} 
 \int\limits^{\gamma}  {\rm d}\gamma' \biggr)     \nonumber \\
&\sim&  \exp \biggl(-\, {1\over{\epsilon \beta_A}}\, {\rho_0\over{\lambda_*}} \,\gamma \biggr)  = 
 \exp \biggl(-\, {\gamma\over\gamma_c}\biggr)   \, ,
\end{eqnarray}
where the exponential cutoff $\gamma_c$ is given by 
\beq
\gamma_c \equiv \epsilon \beta_A {\lambda_*\over{\rho_0}}  = 
\epsilon \beta_A {\lambda_*\over{w_*}} \, \gamma_* \, .
\eeq

This simple example illustrates the idea that (at least in 2D) large plasmoids can serve as particle traps, effectively taking the particles out of the rapid acceleration process, at least temporarily \cite[see][]{Dahlin_etal-2015, Werner_etal-2016, Dahlin_etal-2017, Kagan_etal-2018}.
The origin of the exponential cutoff can then be understood as follows.  To get to a certain high energy, a particle has to spend proportionally longer time in the acceleration region, while all the while being subject to a certain, constant in this case, probability per unit time of being trapped and sequestered in a plasmoid, 
$\tau_{\rm trap}^{-1}(\gamma<\gamma_*) \sim c/\lambda_*$.

In the next two subsections we shall discuss two more realistic examples of~$N(w)$, characterized by broad, power-law distributions.


\subsection{Hierarchical Plasmoid Chain I: a Single Power Law Model}
\label{subsec-plasmoid_chain-simple}

Let us now consider the case when the plasmoid distribution function in the chain is given by a single power law,
\beq
F(w) \sim w^{-\alpha} \, .
\eeq
We normally expect $\alpha \in [1,2]$ in realistic reconnecting plasmoid chains \citep{Uzdensky_etal-2010, Loureiro_etal-2012, Huang_Bhattacharjee-2012, Sironi_etal-2016, Petropoulou_etal-2018}, but here we shall keep the discussion general.

The cumulative distribution $N(w)$ is given by 
\begin{eqnarray}
N(w) &\sim&  {\rm const} - \ln w \qquad {\rm if} \ \alpha=1 \, , \\
N(w) &\sim&  w^{1-\alpha} \qquad \qquad \quad  {\rm if} \ \alpha> 1 \, .
\end{eqnarray}

Let us now further imagine that this power-law distribution extends up to some maximum plasmoid size~$w_{\rm max}$, which we will tentatively associate with the size of the so-called ``monster plasmoids" \citep{Uzdensky_etal-2010, Loureiro_etal-2012}, typically of order $\epsilon L \sim 0.1 L$  \citep{Uzdensky_etal-2010, Loureiro_etal-2012, Sironi_etal-2016, Petropoulou_etal-2018}. 
Then, we can establish the normalization of the plasmoid distribution function by demanding that the number of these largest plasmoids of maximum size~$w_{\rm max}$ found in the chain at any given time is about~1.  
In addition, we should account for the finite length $L$ of the layer, which imposes a strict limit on a particle's free accelerating motion set by the ejection time when a given particle traverses the entire layer, 
$\tau_{\rm ej} \sim c/L$. 
This can be effectively taken into account by requiring that $N$ cannot be less than~1, so that $\lambda_{\rm pl} \leq L$. 
Thus, we shall write
\begin{eqnarray}
N(w \leq w_{\rm max}) &\simeq& 1 - \ln\biggl({w\over w_{\rm max}}\biggr)\, , \quad \alpha = 1 \, , 
\label{eq-N_w-alpha-1} \\
N(w \leq w_{\rm max}) &\simeq& \biggl({w\over w_{\rm max}}\biggr)^{1-\alpha}\, ,  \quad \alpha > 1  \, 
\label{eq-N_w-alpha>1}
\end{eqnarray}

Now, let us consider the interaction of energetic relativistic particles with such a plasmoid chain, using the trapping condition $w \gtrsim w_{\rm tr}(\gamma) = \rho_L(\gamma, B_0) = \gamma \rho_0$ as discussed above. 
It is convenient to define the maximum particle energy that can be confined by the largest plasmoids of size $w_{\rm max}$, i.e., 
\beq
\gamma_{\rm max} \equiv w_{\rm max}/\rho_0 \, .
\eeq

We can then express the inter-plasmoid separation $\lambda_{\rm pl}(w_{\rm tr})$, and hence the typical path-length that a particle can travel before it is captured, as follows.
First, in the special case $\alpha=1$ equation~(\ref{eq-N_w-alpha-1}) yields
\beq
\lambda_{\rm pl}^{\alpha=1}[w_{\rm tr}(\gamma)] = {L\over{N[w_{\rm tr}(\gamma)]}} 
\simeq {L \over{1+ \ln [w_{\rm max}/w_{\rm tr}(\gamma)]}}  =
{L \over{1+ \ln (\gamma_{\rm max}/\gamma)}} \, ,
\label{eq-lambda_pl-a1}
\eeq
which for $\gamma \ll \gamma_{\rm max}$ (and hence for $w_{\rm tr} \ll w_{\rm max}$) can be approximated as 
$\lambda_{\rm pl}^{\alpha=1}[w_{\rm tr}(\gamma \ll \gamma_{\rm max})] \simeq L /\ln (\gamma_{\rm max}/\gamma)$.

Next, for $\alpha > 1$, we have
\beq
\lambda_{\rm pl}[w_{\rm tr}(\gamma)] =  {L\over{N[w_{\rm tr}(\gamma)]}} \simeq 
L \biggl({w_{\rm tr}(\gamma)\over w_{\rm max}}\biggr)^{\alpha-1} \simeq 
L \,  \biggl({\gamma\over{\gamma_{\rm max}}}\biggr)^{\alpha-1} \, , \quad   \alpha > 1\,.
\label{eq-lambda_pl}
\eeq
In particular, in the important special case $\alpha=2$, we have $N(w) \sim w_{\rm max}/w$ and consequently
\beq
\lambda_{\rm pl}^{\alpha=2}[w_{\rm tr}(\gamma)] \simeq  L\, {w_{\rm tr}(\gamma)\over{w_{\rm max}}} = 
L \, {\gamma\over{\gamma_{\rm max}}} = \rho_L(\gamma)\, {L\over w_{\rm max}} \, .
\eeq 
For example, adopting $w_{\rm max} \simeq 0.1 L$, we can estimate this as 
\beq
\lambda_{\rm pl}^{\alpha=2}[w_{\rm tr}(\gamma)] \simeq 10\,w_{\rm tr}(\gamma) \simeq 10\,\rho_L(\gamma) \, .
\eeq

These expressions allow us to estimate the corresponding trapping rates (for $\gamma<\gamma_{\rm max}$):
\beq
\tau_{\rm trap}^{-1}(\gamma)  \sim  
{c\over L}\, \biggl[1 - \ln\biggl({\gamma\over \gamma_{\rm max}}\biggr) \biggr]\, , \quad \alpha = 1\, ,
\eeq
approaching $(c/L) \ln(\gamma_{\rm max}/\gamma)$ for $\gamma \ll \gamma_{\rm max}$;
and
\beq
\tau_{\rm trap}^{-1}(\gamma)  \sim 
{c\over L} N[w(\gamma)] \sim  {c\over L}\, \biggl({\gamma\over{\gamma_{\rm max}}}\biggr)^{1-\alpha}\, , 
\quad \alpha > 1\, , 
\label{eq-tau_trap_alpha>1}
\eeq
and, in particular, 
\beq
\tau_{\rm trap}^{-1}(\gamma)  \sim 
 {c\over L}\, {\gamma_{\rm max}\over{\gamma}} \, ,  \quad \alpha= 2\, . 
\label{eq-tau_trap_alpha=2}
\eeq

Let us now investigate the implications of these estimates for our model of particle acceleration. 
Substituting them into~(\ref{eq-trap}), we find the corresponding expressions for the suppression factors (for $\gamma<\gamma_{\rm max}$):

\smallskip
\noindent \underline{{\bf (i)} $\alpha=1$:}
\beq
\exp \biggl( -\, {\rho_0\over{L\beta_A \epsilon}} \, 
\gamma\,  [2-\ln(\gamma/\gamma_{\rm max})] \biggr) \, .
\eeq

\smallskip
\noindent \underline{{\bf (ii)} $1< \alpha < 2$:}
\beq
\exp \biggl[-\, {w_{\rm max}\over{\epsilon \beta_A L(2-\alpha)}} \, \biggl({\gamma \over{\gamma_{\rm max}}}\biggr)^{2-\alpha}\biggr] = 
\exp \biggl[-\, \biggl({\gamma \over{\gamma_c}}\biggr)^{2-\alpha}\biggr]\, ,
\label{eq-trap_factor-1<alpha<2}
\eeq
where 
\beq
\gamma_c = \gamma_{\rm max} \, \biggl[ \beta_A {{\epsilon L}\over{w_{\rm max}}}\,(2-\alpha) \biggr]^{1\over{2-\alpha}} \, .
\label{eq-gamma_c-1<alpha<2}
\eeq

Interestingly, for $\alpha \rightarrow 1$ this factor approaches a simple exponential cutoff: 
\beq
\exp \biggl[-{\rho_0\over{L}} {1\over{\beta_A \epsilon}} \, \gamma \biggr]  = 
\exp (-\, \gamma/\gamma_{c} )  \, ,
\eeq
where the cutoff Lorentz factor is
\beq
\gamma_{c} \equiv \beta_A \epsilon\, {L\over{\rho_0}} \ , 
\eeq
corresponding to $\rho(\gamma_c) = \gamma_c \rho_0 \simeq \beta_A \epsilon L$ and formally independent of~$w_{\rm max}$.
This case corresponds to a situation where the plasmoid distribution function is so shallow that small plasmoids are not sufficiently numerous to have a strong effect on particle acceleration. The cutoff $\gamma_c$ then simply corresponds to the voltage drop due to the electric field 
$E_{\rm rec} = \epsilon \beta_A B_0$ over the entire layer's length~$L$. 
This limit is thus equivalent to the natural high-energy cutoff expected in small laminar reconnection layers not succumbing to secondary tearing and plasmoid formation~\cite{Larrabee_etal-2003}, as has been confirmed in numerical PIC simulations \cite[e.g.,][]{Lyubarsky_Liverts-2008, Werner_etal-2016}.
It is interesting to note that, if $w_{\rm max} \simeq \epsilon L \simeq 0.1L$, then we can express the cutoff as 
\beq
\gamma_c  \simeq \beta_A \, {w_{\rm max}\over{\rho_0}} = \beta_A \, \gamma_{\rm max}\, . 
\eeq
Thus, in the case of ultra-relativistic reconnection, $\sigma_h \gg 1$ and $\beta_A \rightarrow 1$, this cutoff energy $\gamma_c$ is simply equal to~$\gamma_{\rm max}$, the maximum energy of particles that can be confined in the largest plasmoid (consistent, e.g., with numerical observations by \citealt{Sironi_etal-2016}).  However, in the case of nonrelativistic reconnection, $\beta_A \sim \sigma_h^{1/2} \ll 1$, the cutoff may in principle be smaller than~$\gamma_{\rm max}$.  
Thus, in this case there may be a substantial, measurable range of particle energies where the energy distribution is exponentially suppressed. 

\smallskip
\noindent \underline{{\bf (iii)} $\alpha=2$:} \\
This case is indeed special. There is no exponential cutoff in this case;  instead, particle trapping in plasmoids leads to an additional power-law factor, $\gamma^{-p_{\rm tr}}$, similar to the effect of magnetization by $B_1$ (see \S~\ref{subsec-magnetization}). The corresponding power-law index is given by 
\beq
p_{\rm tr} \simeq {1\over{\epsilon \beta_A}} {w_{\rm max}\over L} \, .
\label{eq-p_tr}
\eeq
This expression is similar to that for the power-law index $p_{\rm magn}$ due to magnetization, see equation~(\ref{eq-p-magn}).  The combined effect of the two processes leads to a power law 
$\gamma^{-p_{\rm tot}^{\alpha=2}}$ with an index given by the sum of the two: 
\beq
p_{\rm tot}^{\alpha=2} = p_{\rm magn}  + p_{\rm tr} \sim 
\beta_A^{-1}\, \biggl(1 + {w_{\rm max}\over{\epsilon L}} \biggr)\, .
\label{eq-p_tot-alpha=2}
\eeq
That is, trapping in plasmoids in the $\alpha=2$ case leads to a steepening of the nonthermal power law compared to expected from magnetization alone, while preserving (in contrast to the $\alpha < 2$ case) the overall power-law shape of the distribution function. 

The relative importance of the magnetization and trapping processes for $\alpha=2$ does not depend on~$\gamma$ and is instead controlled by the ratio~$w_{\rm max}/\epsilon L$. 
In particular, if the chain is truncated at relatively small sizes, i.e., if  $w_{\rm max} \ll \epsilon L$, then $p_{\rm tr} \ll p_{\rm magn}$ and  the trapping correction to the total power-law index is relatively minor. 
If, however, the plasmoid chain extends all the way up to nearly the monster-plasmoid scale, $w_{\rm max} \sim \epsilon L \sim 0.1 L$, then we get 
\beq
p_{\rm tr} \simeq \beta_A^{-1} \sim p_{\rm magn} \, ,
\eeq
and hence the two processes generally play comparable roles for $\gamma< \gamma_{\rm max}$. 
The total power-law index is then greater by a factor of order unity (e.g., double) than that due to magnetization alone, i.e., the resulting steeping is significant. 
We will see, however, in \S~\ref{subsec-Fermi} that Fermi acceleration by moving plasmoids tends to compensate this steepening tendency to some degree and may even cancel it completely.

\subsection{More realistic double-power-law plasmoid chain}
\label{subsec-real-chain}

\begin{figure}
\begin{center}
\includegraphics[width=4 in]{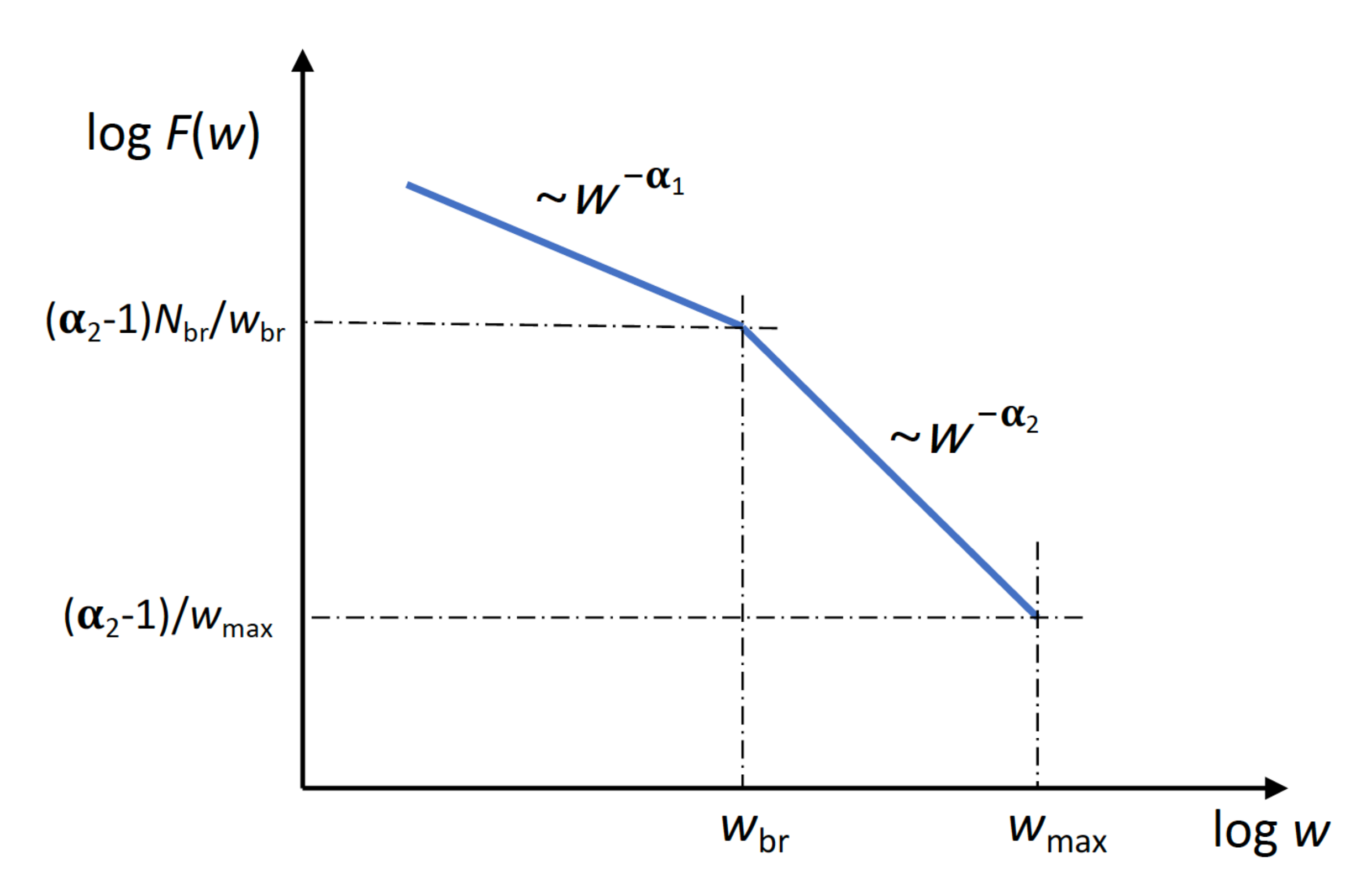}
\end{center}
\caption{A more realistic, broken power-law plasmoid size distribution with a shallow small-$w$ power law $F\propto w^{-\alpha_1}$ transitioning at $w=w_{\rm br}$ to a steeper large-$w$ power law $F\propto w^{-\alpha_2}$, with $\alpha_2>\alpha_1$.}
\label{fig-4}
\end{figure}

Numerical simulations of 2D reconnection in the large-system, plasmoid-dominated regime, done both in the non-relativistic resistive-MHD case \citep{Loureiro_etal-2012, Huang_Bhattacharjee-2012} and in the relativistic collisionless pair-plasma case~\cite{Sironi_etal-2016} (see also \citealt{Petropoulou_etal-2018} for an associated semi-analytical model), show that realistic reconnecting systems develop plasmoid-size distributions that are more complex than a single power law discussed in the previous section.  Namely, the resulting plasmoid distributions are best described by a double power law, i.e., a relatively shallow power law with some index $\alpha_1$ below a certain break plasmoid size $w_{\rm br}$ and another, steeper power law with an index $\alpha_2>\alpha_1$ above~$w_{\rm br}$:
\begin{eqnarray}
F &\sim& w^{-\alpha_1}, \quad w \leq w_{\rm br}, \\
F &\sim& w^{-\alpha_2},  \quad  w_{\rm br} < w \leq w_{\rm max}, 
\end{eqnarray}
with a sharp, perhaps exponential, cutoff above~$w_{\rm max}$ (see figure~\ref{fig-4}). 
The large-$w$ cutoff $w_{\rm max}$ of the second power law may correspond to the monster plasmoid size, typically~$0.1 L$, as in~\S~\ref{subsec-plasmoid_chain-simple}.
Also, based on the above-mentioned simulation studies and analytical theory \citep{Uzdensky_etal-2010}, we generally expect $\alpha_1 \sim 1$ and $\alpha_2 \sim 2$, but here will consider these indices as free variable parameters. 


The intermediate break scale $w_{\rm br}$ that marks the transition between the two power laws will play an important role in our analysis. Unfortunately, what governs it and how it depends on the system parameters is not yet understood; in particular, it is not known whether this size scales with (i.e., just some finite fraction of) the global size~$L$, or with (is a fixed multiple of) the thickness of the smallest elementary inter-plasmoid layers $\delta \sim \bar{\rho} = \bar{\gamma} \rho_0 \sim \sigma_c \rho_0$ (where the latter estimate assumed that the upstream region is magnetically dominated, so that the average particle energy of the heated plasma in the layer is just a finite fraction of $\sigma_c m_e c^2$). 
Answering these important questions is beyond the scope of this paper and should be obtained by more careful numerical studies, e.g., similar to \cite{Sironi_etal-2016}, in conjunction with analytical and semi-analytical studies~\citep[e.g.,][]{Petropoulou_etal-2018}. The analysis presented in this paper provides a strong motivation for such studies as it underscores the strong connection between the statistical properties of the plasmoid chain and~NTPA.  
In the meantime, in this study we will consider $w_{\rm br}$ to be a variable parameter.


For $w \in (w_{\rm br}, w_{\rm max})$, the cumulative plasmoid distribution function is 
\beq 
N (w_{\rm br} < w \leq w_{\rm max}) \sim w^{1-\alpha_1} \, .
\eeq
Using the normalization condition $N(w_{\rm max}) \simeq 1$, we can write 
\beq
N (w_{\rm br} < w \leq w_{\rm max}) \simeq \biggl({w\over{w_{\rm max}}} \biggr)^{1-\alpha_2} \, ,
\label{eq-N_w>wbr}
\eeq
thus also fixing the normalization for $F(w)$ in this range: 
\beq
F (w_{\rm br} < w \leq w_{\rm max}) = -\,N'(w_{\rm br} < w \leq w_{\rm max}) \simeq 
{{\alpha_2-1}\over{w_{\rm max}}} \, \biggl({w\over{w_{\rm max}}} \biggr)^{-\alpha_2} \, .
\label{eq-F_w>wbr}
\eeq

We will also introduce 
\beq
N_{\rm br} \equiv N (w=w_{\rm br}) \simeq \biggl({w_{\rm max}\over{w_{\rm br}}} \biggr)^{\alpha_2-1} \, ,
\label{eq-N_br}
\eeq
\beq
\lambda_{\rm br} \equiv \lambda_{\rm pl}(w_{\rm br}) \equiv  {L\over{N_{\rm br}}}  \simeq
L \, \biggl({w_{\rm br}\over{w_{\rm max}}} \biggr)^{\alpha_2-1} \, ,
\label{eq-lambda_br}
\eeq
and
\beq
\gamma_{\rm br} \equiv \gamma(w_{\rm br}) = {w_{\rm br}\over{\rho_0}} \, ,
\label{eq-gamma_br}
\eeq
to denote, respectively, the total number of plasmoids larger than the critical break size~$w_{\rm br}$, the characteristic separation between them, and the corresponding particle energy.  

\smallskip
We shall now consider the plasmoid distribution below $w_{\rm br}$, which will allow us to estimate the particle trapping factors for $\gamma \leq \gamma_{\rm br}$. The normalization of the plasmoid distribution function $F(w)$ in this range is determined by the requirement that $F(w)$ be continuous at~$w_{\rm br}$:
\beq
F (w<w_{\rm br}) \simeq 
{{\alpha_2-1}\over{w_{\rm br}}} \, N_{\rm br} \, \biggl({w\over{w_{\rm br}}} \biggr)^{-\alpha_1}
\simeq {{\alpha_2-1}\over{w_{\rm max}}} \, \biggl({w_{\rm br}\over{w_{\rm max}}} \biggr)^{-\alpha_2}\, \biggl({w\over{w_{\rm br}}} \biggr)^{-\alpha_1} \, .
\label{eq-F_w<wbr}
\eeq

The cumulative plasmoid distribution function below $w_{\rm br}$ is then determined by a straightforward integration of $F (w<w_{\rm br})$ with the condition that $N(w)$ be continuous at~$w_{\rm br}$. 


In particular, in the case $\alpha_1=1$ we get 
\begin{eqnarray}
F(w<w_{\rm br}) &\simeq & {{\alpha_2-1}\over{w}} N_{\rm br}; \\
N(w\leq w_{\rm br})  &=& \int\limits_w^{w_{\rm br}} F(w') d w' + N(w_{\rm br}) \simeq 
N_{\rm br}\, [1 - (\alpha_2-1) \ln{w\over{w_{\rm br}}}] ; \\
 \int\limits^{\gamma<\gamma_{br}} N[w_{\rm tr}(\gamma')]\, {\rm d} \gamma' 
 &\simeq& N_{\rm br} \gamma\,  \bigl[\alpha_2 - (\alpha_2-1) \ln{\gamma\over{\gamma_{\rm br}}} \bigr] \, .
\end{eqnarray}

Then, using the formalism developed in the previous two subsections [see equation~(\ref{eq-trap})], we obtain the exponential factor describing the particle trapping in plasmoids, for~$\gamma\leq \gamma_{\rm br}$: 
\begin{eqnarray}
&& \exp \biggl(-\, {1\over{\dot\gamma_{\rm acc}}}\, 
 \int\limits^{\gamma} {{\rm d}\gamma'\over{\tau_{\rm trap}(\gamma')}}\biggr) = 
\exp \biggl[-\, {1\over\beta_A}\, {w_{\rm br}\over{\epsilon L}}\, 
 \int\limits^{\gamma} N[w_{\rm tr}(\gamma')]\, {\rm d} \biggl({\gamma' \over{\gamma_{\rm br}}}\biggr) \biggr] 
 \nonumber \\
 &&\simeq \exp \biggl[-\, {1\over\beta_A}\, {w_{\rm br}\over{\epsilon \lambda_{\rm br}}}\, 
 \biggl({\gamma\over{\gamma_{\rm br}}}\biggr)\,  \bigl[\alpha_2 - (\alpha_2-1) \ln{\gamma\over{\gamma_{\rm br}}} \bigr] \biggr] \, ,
\label{eq-trap_gamma<gamma_br_alpha_1=1}
\end{eqnarray}
where we used $\lambda_{\rm br} = L/N_{\rm br}$.  
Thus, ignoring the logarithmic correction, we see that there is simple exponential cutoff, $\exp(-\gamma/\gamma_c)$,  with a cutoff energy that is of order 
\beq
\gamma_c \sim \gamma_{\rm br} \, \beta_A \, \biggl({{\epsilon \lambda_{\rm br}}\over{w_{\rm br}}} \biggr) =
\epsilon \beta_A \, {{\lambda_{\rm br}}\over{\rho_0}} \, .
\label{eq-gamma_c-alpha_1=1}
\eeq

Next, for the case $\alpha_1>1$, integration of (\ref{eq-F_w<wbr}) yields
\beq
N(w\leq w_{\rm br})  = \int\limits_w^{w_{\rm br}} F(w') d w' + N(w_{\rm br}) \simeq 
N_{\rm br} \, \biggl({{\alpha_2-1}\over{\alpha_1-1}}\biggr)\, 
\biggl[ \biggl({w\over{w_{\rm br}}} \biggr)^{1-\alpha_1} - {{\alpha_2-\alpha_1}\over{\alpha_2-1}} \biggr]\, .
\label{eq-N_w<wbr}
\eeq


Then, using (\ref{eq-N_w<wbr}), we can calculate, for $\gamma\leq \gamma_{\rm br}$, 
\begin{eqnarray}
\int \limits^{\gamma} N[w_{\rm tr}(\gamma')]\, d \gamma' 
&=&  {1\over\rho_0} \int \limits^{w_{\rm tr}(\gamma)} \, N(w') d w' \nonumber \\
&\simeq &
\gamma_{\rm br} N_{\rm br} \biggl({{\alpha_2-1}\over{\alpha_1-1}}\biggr)\, 
\biggl[ {1\over{2-\alpha_1}}\, \biggl({w\over{w_{\rm br}}} \biggr)^{2-\alpha_1} - 
{{\alpha_2-\alpha_1}\over{\alpha_2-1}}\,{w\over{w_{\rm br}}} \biggr]_{w=w_{\rm tr}(\gamma)}     \nonumber \\
&=& \gamma N_{\rm br} \biggl({{\alpha_2-1}\over{\alpha_1-1}}\biggr)  
\biggl[ {1\over{2-\alpha_1}} \biggl({\gamma\over{\gamma_{\rm br}}} \biggr)^{1-\alpha_1} - 
{{\alpha_2-\alpha_1}\over{\alpha_2-1}}\biggr]                     \nonumber \\
&=&
\gamma N_{\rm br} \, \biggl[ {{\alpha_2-1}\over{(\alpha_1-1)(2-\alpha_1)}} \, \biggl({\gamma\over{\gamma_{\rm br}}} \biggr)^{1-\alpha_1} - {{\alpha_2-\alpha_1}\over{\alpha_1-1}}\biggr] \, ,
\end{eqnarray}
and thus obtain the particle-trapping factor for~$\gamma\leq \gamma_{\rm br}$: 
\begin{eqnarray}
&& \exp \biggl(-\, {1\over{\dot\gamma_{\rm acc}}}\, 
 \int\limits^{\gamma} {{\rm d}\gamma'\over{\tau_{\rm trap}(\gamma')}}\biggr) = 
 \exp \biggl(-\, {1\over{\epsilon \beta_A \Omega_0}}\, {c\over L} 
 \int\limits^{\gamma}  N[w_{\rm tr}(\gamma')] {\rm d}\gamma' \biggr)  \nonumber \\
&&\simeq \exp \biggl(-\, {1\over{\beta_A}}\, {w_{\rm br}\over{\epsilon \lambda_{\rm br}}} \, 
{\gamma\over{\gamma_{\rm br}}}\, 
\biggl[ {{\alpha_2-1}\over{(\alpha_1-1)(2-\alpha_1)}} \, \biggl({\gamma\over{\gamma_{\rm br}}} \biggr)^{1-\alpha_1} - {{\alpha_2-\alpha_1}\over{\alpha_1-1}}\biggr] \biggr)  \, .
\label{eq-trap_gamma<gamma_br}
\end{eqnarray}

We see that there are two cutoff factors here: 
one pure exponential, $\sim \exp(-\gamma/\gamma_{c1})$, with 
\beq
\gamma_{c1} \simeq  \gamma_{\rm br} \, \beta_A {{\alpha_1-1}\over{\alpha_2-\alpha_1}}  \, 
\biggl({{\epsilon \lambda_{\rm br}}\over{w_{\rm br}}}\biggr) =
\epsilon \beta_A \, {{\alpha_1-1}\over{\alpha_2-\alpha_1}} {{\lambda_{\rm br}}\over{\rho_0}} 
\label{eq-gamma_c1}
\eeq
(similar to what we have obtained above for the $\alpha_1=1$ case, apart from the logarithmic correction and factors of order unity), 
and the other sub-exponential, $\sim \exp[-(\gamma/\gamma_{c2})^{2-\alpha_1}]$, with 
\beq
\gamma_{c2} \simeq  \gamma_{\rm br}\, \biggl[\beta_A {{(\alpha_1-1)(2-\alpha_1)}\over{\alpha_2-1}} \, 
\biggl({{\epsilon \lambda_{\rm br}}\over{w_{\rm br}}}\biggr) \biggr]^{1\over{2-\alpha_1}} \, . 
\label{eq-gamma_c2}
\eeq

As we can see, there is an important factor $\epsilon \lambda_{\rm br}/w_{\rm br}$ that governs the ratio of the cutoff to~$\gamma_{\rm br}$ in all these expressions for any~$\alpha_1$. Assuming first that $\alpha_2<2$, this factor can be evaluated, using~(\ref{eq-lambda_br}), as 
\beq
{{\epsilon \lambda_{\rm br}}\over{w_{\rm br}}} \simeq 
{{\epsilon L}\over{w_{\rm max}}}\,  \biggl({w_{\rm max}\over{w_{\rm br}}}\biggr)^{2-\alpha_2} = 
{{\epsilon L}\over{w_{\rm max}}}\, N_{\rm br}^{{2-\alpha_2}\over{\alpha_2-1}} \, ,
\label{eq-eps-lambda_br/w_br}
\eeq
which is large if $w_{\rm max}\gg w_{\rm br}$ (and hence $N_{\rm br}\gg 1$).
Substituting this into~(\ref{eq-gamma_c1})-(\ref{eq-gamma_c2}), we get
\beq
\gamma_{c1} \simeq  \gamma_{\rm br} \, \beta_A {{\alpha_1-1}\over{\alpha_2-\alpha_1}} \, 
\biggl({{\epsilon L}\over{w_{\rm max}}}\biggr)\,  \biggl({w_{\rm max}\over{w_{\rm br}}}\biggr)^{2-\alpha_2} \, ,
\label{eq-gamma_c1-v2}
\eeq
and
\beq
\gamma_{c2} \simeq  \gamma_{\rm br}\, \biggl[\beta_A {{(\alpha_1-1)(2-\alpha_1)}\over{\alpha_2-1}} \, 
\biggl({{\epsilon L}\over{w_{\rm max}}}\biggr)\, \biggl({w_{\rm max}\over{w_{\rm br}}}\biggr)^{2-\alpha_2}  \biggr]^{1\over{2-\alpha_1}} \, . 
\label{eq-gamma_c2-v2}
\eeq

Thus we can see that in the case of relativistic reconnection, $\beta_A = \mathcal{O}(1)$, the cutoffs 
are large compared to~$\gamma_{\rm br}$ if $\alpha_2<2$; this means that particle trapping in plasmoids is not important for $\gamma < \gamma_{\rm br}$. 
For nonrelativistic ($\beta_A \sim \sigma_h^{1/2} \ll 1$) reconnection, however, particle trapping may become important below $\gamma_{\rm br}$ (i.e., one of the cutoffs may drop below $\gamma_{\rm br}$) if the scale separation between $w_{\rm max}$ and $w_{\rm br}$ is not too large. 
Provided that the numerical coefficients $(\alpha_1-1)$, $(2-\alpha_1)$, $(\alpha_2-\alpha_1)$, etc., are not close to zero,  we can regard them as finite constants of order unity; ignoring them for simplicity, we can then write the condition $\gamma_{c1,2} < \gamma_{\rm br}$ as 
\beq
\beta_A \, {{\epsilon L}\over{w_{\rm max}}} \, 
\biggl({w_{\rm max}\over{w_{\rm br}}}\biggr)^{2-\alpha_2} < 1\, .
\label{eq-cond_gamma_c<gamma_br}
\eeq
If this condition is satisfied, we have $\gamma_{c2} < \gamma_{c1} < \gamma_{\rm br}$, i.e., the cutoff is dominated by~$\gamma_{c2}$. 

\smallskip
Next, we will investigate the relative importance of particle trapping above the break energy~$\gamma_{\rm br}$. 
To do this, we need to evaluate the integral in~(\ref{eq-trap}) for $\gamma >\gamma_{\rm br}$. We will do this by splitting the whole region of integration into two sub-regions: up to~$\gamma_{\rm br}$ and from $\gamma_{\rm br}$ to~$\gamma$, and correspondingly obtain two multiplicative factors. 

The first factor is a constant (independent of~$\gamma$) that can be evaluated as 
\begin{eqnarray}
\exp \biggl(-{c\over{\epsilon \beta_A \Omega_0 L}}  
 \int\limits^{\gamma_{\rm br}}  N[w_{\rm tr}(\gamma')] {\rm d}\gamma' \biggr)  &\simeq& 
 \exp \biggl[-\, {\alpha_2 \over\beta_A}\, {w_{\rm br}\over{\epsilon \lambda_{\rm br}}}\biggr] 
 \qquad \qquad \qquad {\rm for} \ \alpha_1=1 \, , 
 \label{eq-trap_gamma_br_alpha1=1} \\
\exp \biggl(-{c\over{\epsilon \beta_A \Omega_0 L}}  
 \int\limits^{\gamma_{\rm br}}  N[w_{\rm tr}(\gamma')] {\rm d}\gamma' \biggr) &\simeq& 
\exp \biggl(-{1\over{\beta_A}}{w_{\rm br}\over{\epsilon \lambda_{\rm br}}} 
{{1+\alpha_2-\alpha_1}\over{2-\alpha_1}}\biggr)  \  {\rm for} \ \alpha_1>1.
\label{eq-trap_gamma_br_alpha1>1}
\end{eqnarray}
Once again, in light of (\ref{eq-eps-lambda_br/w_br}), and adopting the assumptions $w_{\rm max}/\epsilon L = \mathcal{O}(1)$,  $w_{\rm max}\gg w_{\rm br}$, and $\alpha_2<2$, we see that in both cases this exponential factor is unimportant for relativistic reconnection ($\beta_A \sim 1$). 
But it does become important for extremely non-relativistic reconnection, namely, when 
$\beta_A \simeq \sigma_h^{1/2}$ becomes comparable to $(w_{\rm br}/w_{\rm max})^{2-\alpha_2}$, 
see equation~(\ref{eq-cond_gamma_c<gamma_br}).

To evaluate the second factor, involving the integral of $\tau_{\rm tr}^{-1}(\gamma')$ from $\gamma'=\gamma_{\rm br}$  to $\gamma>\gamma_{\rm br}$, we combine (\ref{eq-tau_trap}) and~(\ref{eq-N_w>wbr})  to get 
\beq
\tau_{\rm tr}^{-1}(\gamma_{\rm br} < \gamma \leq \gamma_{\rm max}) \simeq 
{c\over L} N(w_{\rm br} < w \leq w_{\rm max})|_{w=w_{\rm tr}(\gamma)} \simeq
{c\over L} \biggl({\gamma\over{\gamma_{\rm max}}} \biggr)^{1-\alpha_2} \simeq
{c\over L} N_{\rm br} \biggl({\gamma\over{\gamma_{\rm br}}} \biggr)^{1-\alpha_2},
\eeq
where $\gamma_{\rm max} = w_{\rm max}/\rho_0$. 

The resulting particle trapping factor is then naturally independent of $\alpha_1$ and, for $\alpha_2<2$, is given by 
\beq
\exp \biggl(-\, {1\over{\epsilon \beta_A \Omega_0}}\, {c\over L} 
 \int\limits_{\gamma_{\rm br}}^{\gamma}  N[w_{\rm tr}(\gamma')] {\rm d}\gamma' \biggr) \simeq
\exp \biggl[-\, {w_{\rm max}\over{\beta_A \epsilon L}} \, {1\over{2-\alpha_2}}\, 
\biggl({\gamma\over{\gamma_{\rm max}}}\biggr)^{2-\alpha_2} \biggr] \, ,
\eeq
which is, of course, identical to (\ref{eq-trap_factor-1<alpha<2}) obtained in the single power-law plasmoid distribution case. The resulting cutoff energy 
\beq
\gamma_c = \gamma_{\rm max} \, \biggl[ \beta_A {{\epsilon L}\over{w_{\rm max}}}\,(2-\alpha_2) \biggr]^{1\over{2-\alpha_2}} 
\label{eq-gamma_c}
\eeq
can also be expressed in terms of $\gamma_{\rm br}$ as
\beq
\gamma_c = 
\gamma_{\rm br} \, \biggl[ \beta_A {{\epsilon L}\over{w_{\rm max}}}\,(2-\alpha_2) \biggl({w_{\rm max}\over{w_{\rm br}}}\biggr)^{2-\alpha_2} \biggr]^{1\over{2-\alpha_2}} \,.
\label{eq-gamma_c-v2}
\eeq
We then see that $\gamma_c$ is above $\gamma_{\rm br}$ unless the inequality~(\ref{eq-cond_gamma_c<gamma_br}) is satisfied. If, however, the extreme nonrelativistic condition~(\ref{eq-cond_gamma_c<gamma_br}) is satisfied, then $\gamma_c$ formally drops below~$\gamma_{\rm br}$ [and even below~$\gamma_{c1}$ and $\gamma_{c2}$, as one can see by by comparing (\ref{eq-gamma_c-v2}) with~(\ref{eq-gamma_c1-v2})-\ref{eq-gamma_c2-v2}) and ignoring order-unity factors like $(2-\alpha_2)$]. This means that the particle distribution suffers even more severe suppression above $\gamma_{\rm br}$ in this case.
%

\smallskip
The case $\alpha_2=2$ is, again, special.  Examining this case is motivated by analytical arguments \citep{Uzdensky_etal-2010} and numerical evidence \citep{Loureiro_etal-2012, Sironi_etal-2016} that this situation is indeed realized in practice. 
In this case, $\lambda_{\rm br} \simeq w_{\rm br} L/w_{\rm max}$ [see~(\ref{eq-lambda_br})], and hence 
\beq
{{\epsilon \lambda_{\rm br}}\over{w_{\rm br}}} \simeq {{\epsilon L}\over{w_{\rm max}}}\, . 
\eeq
Thus, since we generally expect $w_{\rm max} \sim \epsilon L$, the cutoff of the small-$\gamma$ ($\gamma < \gamma_{\rm br}$) segment of the particle distribution 
[see~(\ref{eq-gamma_c-alpha_1=1}) and (\ref{eq-gamma_c1})-(\ref{eq-gamma_c2})] 
can be estimated, ignoring factors of order unity and the logarithmic correction in the $\alpha_1=1$ case, as 
\beq
\gamma_c \sim \gamma_{\rm br} \, \beta_A \, {{\epsilon \lambda_{\rm br}}\over{w_{\rm br}}} \sim
\gamma_{\rm br} \, \beta_A \, {{\epsilon L}\over{w_{\rm max}}} \sim 
\gamma_{\rm br} \, \beta_A \, .
\label{eq-gamma_c-alpha_2=2}
\eeq
That is, the cutoff is comparable to $\gamma_{\rm br}$ for relativistic reconnection, and small compared to $\gamma_{\rm br}$ for non-relativistic reconnection. This indicates that the trapping of particles by plasmoids is important in this case even for moderate-energy ($\gamma \lesssim \gamma_{\rm br}$) relativistic particles. 
In particular, in the ultra-relativistic reconnection case, $\beta_A \simeq 1$, we get 
$\gamma_c \sim \gamma_{\rm br} = w_{\rm br}/\rho_0$, which can be recast as 
$\gamma_c \sim  \sigma_c w_{\rm br}/\rho_{\sigma}$, where $\rho_{\sigma} \equiv \sigma_c \rho_0$. 
We then see that if $w_{\rm br}$ scales as a finite multiple of $\delta \sim \rho_{\sigma}$, then we get a result consistent with \cite{Werner_etal-2016}, i.e., the existence of an exponential cutoff proportional to~$\sigma_c$, e.g., $\gamma_{c1} \simeq 4 \sigma_c$. 

Furthermore, for particles above the break energy $\gamma_{\rm br}$, the trapping by large plasmoids manifests itself not via an exponential or quasi-exponential suppression factor but, just as in the case of a single power-law plasmoid chain with $\alpha=2$ [see~(\ref{eq-p_tr})], via a power-law steepening due to an additional power-law factor: 
\beq
\exp \biggl(-\, {1\over{\epsilon \beta_A}}\, {\rho_0\over L} 
 \int\limits_{\gamma_{\rm br}}^{\gamma}  N[w_{\rm tr}(\gamma')] {\rm d}\gamma' \biggr) \simeq
\exp \biggl[-\, {w_{\rm max}\over{\beta_A\epsilon L}}\, \ln\biggl({\gamma\over{\gamma_{\rm br}}}\biggr) \biggr] 
\simeq \biggl({\gamma\over{\gamma_{\rm br}}}\biggr)^{-p_{\rm tr}}  \, ,
\eeq
with the same power-law index as in~(\ref{eq-p_tr}):
\beq
p_{\rm tr} \simeq \beta_A^{-1}\, {w_{\rm max}\over{\epsilon L}} \, .
\eeq
Once again, assuming that $w_{\rm max} \sim \epsilon L$, we see that this additional power-law index resulting from particle trapping in plasmoids scales as 
$p_{\rm tr} \sim \beta_A^{-1} \sim (1+1/\sigma_h)^{1/2}$.

\smallskip
To sum up, for $\alpha_2 < 2$ and relativistic reconnection ($\beta_A \sim 1$) we find that the existence of the power-law break in the plasmoid distribution at $w_{\rm br} \ll w_{\rm max}$ does not result in any significant changes to the particle energy distribution relative to the case of a single power-law with $\alpha = \alpha_2$. 
However, in the case of nonrelativistic reconnection ($\beta_A \ll 1$), the presence of the break in the plasmoid spectrum $F(w)$ may lead to a modification in the particle distribution if the dynamic range $w_{\rm max}/w_{\rm br}$ of the second ($\alpha = \alpha_2$) power-law segment of$F(w)$ is not too large, namely if the condition~(\ref{eq-cond_gamma_c<gamma_br}) is satisfied. 
In this case, the high-energy cutoff of the particle distribution falls below $\gamma_{\rm br}$ and is given by equation~(\ref{eq-gamma_c2-v2}). 

In the special case $\alpha_2 =2$ we expect a power-law distribution, with the usual index $p_{\rm magn}$ scaling inversely with~$\beta_A$, at moderately suprathermal energies ($\gamma \lesssim \gamma_{\rm br}$), checked by an exponential or quasi-exponential cutoff~(\ref{eq-gamma_c-alpha_2=2}). At higher energies, above $\gamma_{\rm br}$, the distribution transitions to a second, steeper power law with index $p_{\rm tot} = p_{\rm magn} + p_{\rm tr}$.
The cutoff $\gamma_c$ connecting the two power laws is of order $\gamma_{\rm br} \beta_A$, and is thus fundamentally controlled by the break $w_{\rm br} = \gamma_{\rm br} \rho_0$ in the plasmoid size distribution; in particular, it can be much smaller than the extreme-acceleration Hillas limit of~$\sim L/\rho_0$, and may even just scale with~$\sigma_c \beta_A$ if $w_{\rm br} \sim \sigma_c \rho_0$. 
Moreover, in the case of strongly non-relativistic reconnection, $\beta_A \sim \sigma_h^{1/2} \ll 1$, the cutoff $\gamma_c$ becomes much smaller than~$\gamma_{\rm br}$, thus greatly reducing the dynamic range of the first power-law segment of the particle distribution. This has a strong effect on the normalization of the second, high-energy power-law segment above~$\gamma_{\rm br}$: its normalization is suppressed by the exponential factors (\ref{eq-trap_gamma_br_alpha1=1})-(\ref{eq-trap_gamma_br_alpha1>1}); the suppression is just by a finite factor of order unity in the ultra-relativistic case $\beta_A \simeq 1$ but could be exponentially strong, {\it viz.} $\exp(-1/\beta_A)$, in the non-relativistic case $\beta_A \ll 1$. This may explain the difficulties in obtaining robust extended nonthermal relativistic power laws in numerical PIC studies of 2D nonrelativistic magnetic reconnection.



\subsection{Fermi acceleration by reflection off moving plasmoids}
\label{subsec-Fermi}

In this subsection we discuss the diffusive Fermi acceleration due to a particle's reflection off moving plasmoids that fly chaotically up and down along the plasmoid chain (see also \citealt{Nalewajko_etal-2015, Guo_etal-2015}). We stress again that this acceleration avenue is distinct from the above-considered acceleration by the main (regular) reconnection electric field $E_{\rm rec}$ responsible for the growth of plasmoids (see~ \S~\ref{subsec-accel}). Indeed, since the Fermi acceleration process under consideration here is related to the relative motions of plasmoids along the layer (i.e., in the $x$-direction),  it is fundamentally driven by the nonlinear development of the coalescence instability, as opposed to the tearing instability associated with~$E_{\rm rec}$. 
Building on the qualitative discussion of the underlying physics of this process in \S~\ref{subsec-picture-particle-acceleration}, here we develop a more quantitative description.

The detailed microphysical underpinning of this acceleration process can be described as follows (see figure~\ref{fig-3}). 
When a particle in its motion in the $(xz)$-plane encounters a large plasmoid moving (relative to the lab frame) with a speed~$v_{\rm pl}$, it interacts with the plasmoid's motional electric field of magnitude $|E_z| \sim E_{\rm pl} = v_{\rm pl} B_{\rm pl} /c$; the sign of this electric field can be positive or negative depending on the plasmoid's direction of motion. 
During this interaction, as discussed in \S~\ref{subsec-picture-particle-acceleration}, the particle completes a $\gtrsim 1/2$ fraction of its cyclotron orbit in the plasmoid's magnetic field, thus moving backward in the $z$-direction by about the orbit's diameter, $|\Delta z|  \sim 2 \gamma  m_e c^2/eB_{\rm pl}$. Thus, the magnitude of the resulting particle's energy change due to the work done on it by the electric field $E_{\rm pl}$ during this interaction can be estimated as
\beq
|\Delta \epsilon| \simeq e E_{\rm pl} |\Delta z| \simeq (v_{\rm pl}/c) \, \gamma m_e c^2  = 
(v_{\rm pl}/c) \, \epsilon \, . 
\eeq
For simplicity, we shall take the characteristic plasmoid velocities to be independent of plasmoid size, namely,  of order $v_{\rm pl} \sim V_A = \beta_A c$ for all plasmoids.%
\footnote{We acknowledge, however, that this assumption may be too simplistic; in reality, big plasmoids may take longer to accelerate to the full Alfv\'en speed and thus may be moving slower than smaller ones, and hence somewhat slower than~$V_A$ \cite[see, e.g.,][for discussion]{Sironi_etal-2016, Petropoulou_etal-2018}. Since higher-energy particles can be reflected only by sufficiently large plasmoids, the $v_{\rm pl}(\gamma)/V_A$ factor may be a decreasing function of particle energy, suppressing the reflection-based Fermi acceleration for higher-energy particles.}
Then, the particle's fractional energy change is of order 
\beq
{|\Delta \epsilon|\over{\epsilon}} = {|\Delta \gamma|\over{\gamma}} \simeq {v_{\rm pl} \over c} \sim \beta_A  \, ,
\eeq
which is consistent with the result of an elastic collision in the plasmoid frame.

As discussed in \S~\ref{subsec-picture-particle-acceleration}, the sign of this energy change can be positive or negative depending on whether the encounter is head-on or tail-on. If we ignore correlations in the neighboring large plasmoids' directions of motion%
\footnote{This assumption is not necessarily justified in the case of a particle trapped between two large plasmoids heading towards each other (and subsequently merging); in this case the particle can experience 1st-order Fermi acceleration.}
and thus regard the scatterings as random, the particle's energy undergoes a random walk, resulting in diffusive second-order Fermi acceleration. 

For a high-energy particle with a Lorentz factor $\gamma \gg \bar{\gamma}$, the characteristic time-step of this random walk, i.e.,  the interval between successive encounters with large plasmoids, is approximately the particle's flight time over the typical inter-plasmoid distance between such large plasmoids, i.e., $\Delta t (\gamma) \sim \lambda_{\rm pl}[w_{\rm tr} (\gamma)]/c $. 
Then, the corresponding energy diffusion coefficient can be estimated, ignoring factors of order unity, as
\beq
D_\gamma(\gamma) = {{\Delta \gamma^2}\over{\Delta t}} \sim 
\beta_A^2  \gamma^2 \, {c\over{\lambda_{\rm pl}[w_{\rm tr}(\gamma)]}} =  
\beta_A^2  \gamma^2 \, {c\over L}\, N_{\rm pl}[w_{\rm tr}(\gamma)] = 
\beta_A^2  \gamma^2 \, \tau_{\rm trap}^{-1}(\gamma) \, , 
\label{eq-D_gamma}
\eeq
and the corresponding term in the kinetic equation~(\ref{eq-kinetic-1}) or~(\ref{eq-kinetic-2}) becomes
\beq
\partial_\gamma (D_\gamma \partial_\gamma f) \sim
\beta_A^2 \, {c\over L}\,  \partial_\gamma \biggl( \gamma^2  N_{\rm pl}[w_{\rm tr}(\gamma)]\, \partial_\gamma f \biggr) =
 \beta_A^2 \,  \partial_\gamma \biggl( \gamma^2  \tau_{\rm trap}^{-1}(\gamma) \partial_\gamma f \biggr) \, .
\label{eq-Fermi_diff_term}
\eeq

\smallskip 
We can get further insight by defining the nominal characteristic diffusive Fermi acceleration time
\beq
\tau_{\rm dif}(\gamma) \equiv {\gamma^2\over{D_\gamma(\gamma)}}  \sim 
\beta_A^{-2}  \, {{\lambda_{\rm pl}[w_{\rm tr}(\gamma)]}\over c} =
\beta_A^{-2}  \, {L\over c} \, {1\over{N_{\rm pl}[w_{\rm tr}(\gamma)]}} \, ,
\label{eq-tau_dif}
\eeq
and comparing it to the characteristic timescales corresponding to the other terms in the kinetic equation~(\ref{eq-kinetic-2}). 

First, the ratio of $\tau_{\rm dif}$ to the particle escape time due to trapping by plasmoids is particularly simple and is independent of the particle energy:
\beq
{\tau_{\rm dif}(\gamma) \over{\tau_{\rm trap}(\gamma)}} \sim \beta_A^{-2} = 1+1/\sigma_h \, .
\label{eq-tau_diff-tau_trap}
\eeq
Thus, the two timescales are automatically comparable in the case of relativistic reconnection, $\sigma_h \gtrsim 1$, $\beta_A \simeq 1$; but for nonrelativistic reconnection, $\sigma_h \ll 1$, we find that $\tau_{\rm dif} \gg  \tau_{\rm trap}$. 

Next, by comparing (\ref{eq-tau_dif}) to the magnetization timescale $\tau_{\rm magn}$ [see equation~(\ref{eq-tau_magn})] and the regular acceleration timescale 
$ \tau_{\rm reg}(\gamma) \equiv \gamma/\dot{\gamma}_{\rm acc} \sim \rho(\gamma)/c \epsilon \beta_A$ [see equation~(\ref{eq-gamma-dot})], we find: 
\beq
{\tau_{\rm dif}(\gamma) \over{\tau_{\rm magn}(\gamma)}} \sim
\beta_A^{-2}  \, {{\lambda_{\rm pl}[w_{\rm tr}(\gamma)]}\over c} {\epsilon\Omega_0\over\gamma} =
\beta_A^{-2}  \, {{\epsilon\lambda_{\rm pl}[w_{\rm tr}(\gamma)]}\over{\rho(\gamma)}} \sim
\beta_A^{-2}\, {{\epsilon\lambda_{\rm pl}(w)}\over{w}} \biggl|_{w=w_{\rm tr}(\gamma)} \, ,
\label{eq-tau_diff-tau_magn-1}
\eeq
and
\beq
{\tau_{\rm dif}(\gamma) \over{\tau_{\rm reg}(\gamma)}} \sim
\beta_A^{-2}  \, {{\lambda_{\rm pl}[w_{\rm tr}(\gamma)]}\over c} {c \epsilon\beta_A \over{\rho(\gamma)}} \simeq
\beta_A^{-1}  \, {{\epsilon\lambda_{\rm pl}[w_{\rm tr}(\gamma)]}\over{\rho(\gamma)}} \sim
\beta_A^{-1}\, {{\epsilon\lambda_{\rm pl}(w)}\over{w}} \biggl|_{w=w_{\rm tr}(\gamma)} \, .
\label{eq-tau_diff-tau_reg-1}
\eeq

We can make two observations from these two expressions. 
First, the last factor in these expressions, $\epsilon \lambda_{\rm pl}(w)/w = \epsilon L/[wN(w)]$, already familiar from the discussion in \S~\ref{subsec-real-chain}, manifestly  underscores that the relative importance of Fermi acceleration (compared to particle magnetization by~$B_1$ and regular acceleration by~$E_{\rm rec}$) depends on the plasmoid distribution function~$N(w)$. 
Thus, further analysis requires adopting a specific choice for~$N(w)$. 
For concreteness, let us consider the case of a single power-law plasmoid distribution~(\ref{eq-N_w-alpha>1}) with~$\alpha>1$ (see \S~\ref{subsec-plasmoid_chain-simple}): 
$N\simeq (w/w_{\rm max})^{1-\alpha}$.  
We then have 
\begin{eqnarray}
D_\gamma(\gamma) &\sim &
\beta_A^2\, \gamma_{\rm max}^2 {c\over L}\, \biggl({\gamma\over{\gamma_{\rm max}}}\biggr)^{3-\alpha},
\label{eq-D_gamma-4power_law} \\
\tau_{\rm dif}(\gamma)&\sim & \beta_A^{-2} \, {L\over c}\, 
\biggl({\gamma\over{\gamma_{\rm max}}}\biggr)^{\alpha-1} \, , 
\end{eqnarray}
and the factor $\epsilon \lambda_{\rm pl}(w)/w$, controlling the ratios of $\tau_{\rm dif}$ to $\tau_{\rm magn}$ and~$\tau_{\rm reg}$, can be written as 
\beq
{{\epsilon\lambda_{\rm pl}(w)}\over{w}} \simeq 
{\epsilon L\over w} \biggl({w\over{w_{\rm max}}}\biggr)^{\alpha-1} = 
 {\epsilon L\over w_{\rm max}} \biggl({w\over{w_{\rm max}}}\biggr)^{\alpha-2} 
\, .
\label{eq-eps_lambda/w}
\eeq

Substituting this estimate into equations~(\ref{eq-tau_diff-tau_magn-1})-(\ref{eq-tau_diff-tau_reg-1}), we obtain
\beq
{\tau_{\rm dif}(\gamma) \over{\tau_{\rm magn}(\gamma)}} \sim
\beta_A^{-2}\, {\epsilon L\over w_{\rm max}} \, \biggl({\gamma\over{\gamma_{\rm max}}}\biggr)^{\alpha-2} \, , 
\label{eq-tau_diff-tau_magn-2}
\eeq
and
\beq
{\tau_{\rm dif}(\gamma) \over{\tau_{\rm reg}(\gamma)}} \sim
\beta_A^{-1}\, {\epsilon L\over w_{\rm max}} \, \biggl({\gamma\over{\gamma_{\rm max}}}\biggr)^{\alpha-2} \, .
\label{eq-tau_diff-tau_reg-2}
\eeq

Considering first the case $\alpha<2$ and adopting a reasonable assumption that $w_{\rm max}\sim \epsilon L \sim L/10$, we see that the ratio $\epsilon \lambda_{\rm pl}(w)/w$ given by (\ref{eq-eps_lambda/w}) can be expected to be of order unity only for the biggest plasmoids, $w\sim w_{\rm max}$.  For the majority of plasmoids, however, $w \ll w_{\rm max}$, and this ratio becomes large. 
Then, since $\beta_A \leq 1$, equations (\ref{eq-tau_diff-tau_magn-2})-(\ref{eq-tau_diff-tau_reg-2}) imply that, for the majority of the nonthermal particles, namely those with $\gamma \ll \gamma_{\rm max}$, the diffusive Fermi acceleration by moving plasmoids is relatively unimportant compared to the regular acceleration by~$E_{\rm rec}$ and magnetization by~$B_1$, and, furthermore, that its relative importance decreases as the particle energy is lowered. In other words, while smaller plasmoids, which are capable of scattering lower-energy particles, are more numerous, they are not numerous enough as long as the plasmoid spectrum is sufficiently shallow, $\alpha < 2$. On the other hand, however, for the highest-energy particles with $\gamma\sim \gamma_{\rm max}$ [corresponding to $w_{\rm tr}(\gamma) \sim w_{\rm max}$] the diffusive Fermi acceleration by moving plasmoids may still be important, provided that the $\beta_A^{-1}$ and $\beta_A^{-2}$ factors can be circumvented (see below).  

Next, let us consider the practically important case $\alpha=2$, which is, again, special. In this case, equation~(\ref{eq-eps_lambda/w}) tells us that $\epsilon\lambda_{\rm pl}/w$ is independent of~$w$, and hence the ratios $\tau_{\rm dif}/\tau_{\rm magn}$ and~$\tau_{\rm dif}/\tau_{\rm reg}$ are the same for all particles in the nonthermal power law, and are just governed by~$\beta_A$. 
For reference, the diffusion coefficient and the nominal diffusive acceleration time in this case are given by
\begin{eqnarray}
D_\gamma^{\alpha=2}(\gamma) &\sim &
\beta_A^2\, \gamma_{\rm max}^2 {c\over L}\, {\gamma\over{\gamma_{\rm max}}},\\
\tau_{\rm dif}^{\alpha=2}(\gamma)&\sim & \beta_A^{-2} \, {L\over c}\, 
{\gamma\over{\gamma_{\rm max}}} \, .
\end{eqnarray}

\smallskip
The second observation we can make by looking at equations (\ref{eq-tau_diff-tau_magn-1})-(\ref{eq-tau_diff-tau_reg-1}), as well as equation~(\ref{eq-tau_diff-tau_trap}), is  that the relative strength of the Fermi acceleration appears to be suppressed in the nonrelativistic ($\beta_A\ll 1$) reconnection case relative to the relativistic ($\beta_A \simeq 1$) case by the factors of $\beta_A$ or~$\beta_A^2$. 
We point out, however, that it is not quite fair to judge the relative importance of Fermi acceleration just by the ratios of $\tau_{\rm dif}$ to the other characteristic timescales in the kinetic equation, especially in the non-relativistic case. The reason for this is that the diffusion term involves higher-order (namely, 2nd-order) derivatives of $f(\gamma)$ than the other terms. Thus, if $f(\gamma)$ has a strong dependence on $\gamma$ (e.g., declines rapidly), then this term may become important (although not dominant). In particular, for a power-law spectrum $f(\gamma) \sim \gamma^{-p}$, corresponding to a power-law plasmoid distribution with an index~$\alpha$, we can estimate the diffusion term (\ref{eq-Fermi_diff_term}), using equation~(\ref{eq-D_gamma-4power_law}), as 
\beq
\partial_\gamma (D_\gamma \partial_\gamma f) \sim
p (\alpha+p-2)\, \beta_A^2 \, {c\over L} \,
 \biggl({\gamma\over{\gamma_{\rm max}}}\biggr)^{1-\alpha} f(\gamma) 
 \sim  p (\alpha+p-2)\, {f(\gamma)\over{\tau_{\rm dif}(\gamma)}} \, ,
\label{eq-Fermi_diff_term-alpha}
\eeq
which thus can be significantly greater than just the naive estimate $f/\tau_{\rm dif}$ if $p \gg 1$.
This is in fact expected in the nonrelativistic reconnection case, where $p \sim \beta_A^{-1} \gg 1$ [see~(\ref{eq-p-magn}) and~(\ref{eq-p_tot-alpha=2})];  in this case, the diffusion term is by a factor $\beta_A^{-2} \simeq \sigma_h^{-1} \gg 1$ greater than $f/\tau_{\rm dif}$ if $p \gg 1$, and becomes automatically comparable to the plasmoid-trapping term.

In the special case $\alpha=2$ expression~(\ref{eq-Fermi_diff_term-alpha}), for any $\beta_A$, simplifies to 
\beq
\partial_\gamma (D_\gamma \partial_\gamma f)|_{\alpha=2} \sim
p^2 \beta_A^2 \, {c\over L} \, {\gamma_{\rm max}\over{\gamma}}\,  f(\gamma) \sim
p^2 {f(\gamma)\over{\tau_{\rm dif}(\gamma)}}\, ,
\label{eq-Fermi_diff_term-alpha2}
\eeq
which is, again, by a factor $p^2 \sim \beta_A^{-2} \sim \sigma_h^{-1}$ greater than $f/\tau_{\rm dif}$ in the nonrelativistic limit.
 
We can then see that the Fermi diffusion term may affect our calculation of the particle spectrum power-law index~$p$. 
To remind the reader, in our theory this index is governed, in the absence of Fermi acceleration, by the interplay between the regular acceleration $\dot{\gamma}_{\rm acc}$ by the main reconnection electric field $E_{\rm rec}$ and particle magnetization by the reconnected field~$B_1$ (see \S\S~\ref{subsec-accel} and~\ref{subsec-magnetization}).  And in the special case $\alpha=2$, particle trapping in plasmoids also affects the power-law index, resulting in a moderate steepening [see \S~\ref{subsec-plasmoid_chain-simple} and, in particular, equation~(\ref{eq-p_tot-alpha=2})]. We can now examine the role of diffusive Fermi acceleration in the balance between all these terms, which controls~$p$.  

To do this, let us consider the steady state kinetic equation~(\ref{eq-kinetic-2}) and focus on the power-law part of the particle distribution function, i.e., $f\sim \gamma^{-p}$, thus ignoring a possible high-energy cutoff due to trapping by large plasmoids. 
Then, considering first the case $1< \alpha<2$, and hence excluding the plasmoid-trapping term, we can write this equation as
\beq
\epsilon \beta_A \Omega_0 p \, {f\over\gamma} -\, \epsilon \Omega_0 \, {f\over\gamma} 
+ p (\alpha+p-2)\, \beta_A^2 \, {c\over L} \,
 \biggl({\gamma\over{\gamma_{\rm max}}}\biggr)^{1-\alpha} f(\gamma) = 0 \, , 
\label{eq-kinetic-Fermi}
\eeq
where we have used equations (\ref{eq-gamma-dot}) and~(\ref{eq-tau_magn}) to evaluate the first two terms (regular acceleration by $E_{\rm rec}$ and magnetization by~$B_1$). 
Cancelling $f$, multiplying through by $\gamma/\epsilon \Omega_0$, and using $w_{\rm max} = \gamma_{\rm max} \rho_0 = \gamma_{\rm max} c \Omega_0^{-1}$, we obtain
\beq
\beta_A p  - 1 + p (\alpha+p-2)\, \beta_A^2 \, {w_{\rm max}\over{\epsilon L}} \,
 \biggl({\gamma\over{\gamma_{\rm max}}}\biggr)^{2-\alpha}  = 0 \, , 
\eeq
Since we here assume $\alpha < 2$, we again see that, with the exception of the most energetic particles with $\gamma\sim \gamma_{\rm max}$, the last term, corresponding to Fermi acceleration by moving large plasmoids, is small --- consistent with our discussion after equation~(\ref{eq-tau_diff-tau_reg-2}) above. We then get a familiar result $p \simeq \beta_A^{-1}$. 

Next let us consider the special case $\alpha=2$. In this case, the problem can be solved exactly. 
The particle trapping by plasmoids now affects the power-law index instead of introducing a high-energy cutoff (see \S~\ref{subsec-plasmoid_chain-simple}), and needs to be included in the above kinetic equation~(\ref{eq-kinetic-Fermi}). Using equation~(\ref{eq-tau_trap_alpha=2}) to evaluate this term, equation~(\ref{eq-kinetic-Fermi}) is modified as 
\beq
\epsilon \beta_A \Omega_0 p \, {f\over\gamma} -\, \epsilon \Omega_0 \, {f\over\gamma} 
- \, {c\over L} \,  \biggl({\gamma_{\rm max}\over{\gamma}}\biggr) \, f(\gamma) 
+ p^2\, \beta_A^2 \, {c\over L} \,
 \biggl({\gamma_{\rm max}\over{\gamma}}\biggr) f(\gamma) = 0 \, , 
\eeq
which simplifies to the following quadratic equation for $p\beta_A$, involving a single dimensionless parameter~$w_{\rm max}/{\epsilon L}$: 
\beq
p  \beta_A - \biggl(1 + \, {w_{\rm max}\over{\epsilon L}}\biggr)
+ p^2  \beta_A^2 \, {w_{\rm max}\over{\epsilon L}}  = 0 \, . 
\eeq
The positive solution of this equation, for any $w_{\rm max}/{\epsilon L}$, is simply 
\beq
p^{\alpha=2} = \beta_A^{-1} \, ,
\label{eq-p-alpha=2_withFermi}
\eeq
coinciding with~(\ref{eq-p-magn}). Thus, in the $\alpha=2$ case, diffusive Fermi acceleration by rapidly moving plasmoids effectively cancels, or negates, the steepening of the particle distribution due to particle trapping in plasmoids described by~(\ref{eq-p_tot-alpha=2}). Note, however, that in reality the diffusion term may not lead  to an exact complete cancellation of the trapping effect because, in a more precise and realistic model, both of these terms would in general come in with some numerical coefficients of order unity, which don't have to be equal. We discuss this more general situation in the next subsection.


\subsection{General Form of the Kinetic Equation}
\label{subsec-general_kin_eq}

Finally, for completeness and to set the stage for future, more detailed studies of this problem, in this subsection we present, for reference, the general 2nd-order linear ordinary differential equation for the stationary particle distribution function including all the terms, for an arbitrary plasmoid distribution. 
To obtain it, we plug in the estimates (\ref{eq-gamma-dot}), (\ref{eq-tau_magn}), and (\ref{eq-Fermi_diff_term}) into the steady-state kinetic equation~(\ref{eq-kinetic-2}).  Note however, that, generally speaking, all these estimates are uncertain up to factors of order unity; we will treat these uncertainties by assigning some unknown positive constant coefficients, $c_1$, $c_2$, $c_3$, to the last three terms when we enter these expressions into~(\ref{eq-kinetic-2}).  We then obtain: 
\beq
 f'(\gamma) + c_1 {1\over{\beta_A}} {f\over\gamma} +
 c_2 {1\over{\beta_A }} {f\over{\epsilon\Omega_0 \tau_{\rm trap}(\gamma)}} -
 c_3 \beta_A \biggl( \gamma^2 {1\over{\epsilon\Omega_0 \tau_{\rm trap}(\gamma)}} f' \biggr)' =0 \, , 
\label{eq-kin-general-1}
\eeq
where prime denotes derivative with respect to~$\gamma$. 

Using $\tau_{\rm trap}(\gamma)  \simeq \lambda_{\rm pl}[w_{\rm tr}(\gamma)]/c$, and 
$c/\Omega_0 = \rho_0 = w_{\rm tr}(\gamma) /\gamma$, it is convenient to define a new function encapsulating the information about the plasmoid distribution, 
\beq
q(\gamma) \equiv {w\over{\epsilon \lambda_{\rm pl}(w)}} \vert_{w=w_{\rm tr}(\gamma) = \gamma \rho_0} ={\gamma\over{\epsilon\Omega_0 \tau_{\rm trap}(\gamma)}} \, .
\eeq
Then equation~(\ref{eq-kin-general-1}) becomes
\beq
 f'(\gamma) +  {1\over{\beta_A}} {f\over\gamma}\, [c_1 + c_2  q(\gamma)] -
c_3 \beta_A \biggl( \gamma q(\gamma) f' \biggr)' =0 \, , 
\label{eq-kin-general-2}
\eeq



Using the substitution $\gamma = e^{\beta_A t}$ and $\tilde{q}(t) = q[\gamma(t)]$, 
we can rewrite this as 
\beq
 \dot{f} (t) +   f(t)\, [c_1 + c_2  \tilde{q}(t)] -
c_3 \, {d\over{dt}} \biggl( \tilde{q}(t) \dot{f} \biggr) =0 \, , 
\label{eq-kin-general-3}
\eeq
where $\dot{f} \equiv df/dt$. 

In particular, if the plasmoid distribution is a single power law with $\alpha>1$, then, using~(\ref{eq-tau_trap_alpha>1}), 
\beq
q(\gamma) = q_0 \biggl({\gamma\over{\gamma_{\rm max}}}\biggr)^{2-\alpha} = 
q_0 \, e^{(2-\alpha)\beta_A (t-t_{\rm max})}
\, , 
\eeq
where we defined $q_0 = w_{\rm max}/\epsilon L$, and 
$t_{\rm max} \equiv \beta_A^{-1}\ln\gamma_{\rm max}$. 
Equation~(\ref{eq-kin-general-3}) then transforms into 
\beq
 \dot{f} (t) +   f(t)\, [c_1 + c_2  q_0 e^{(2-\alpha)\beta_A (t - t_{\rm max})}] -
c_3 \, {d\over{dt}} \biggl( q_0 e^{(2-\alpha)\beta_A (t - t_{\rm max})} \dot{f} \biggr) =0 \, . 
\label{eq-kin-general-4}
\eeq

For example, in the special case $\alpha = 2$, we get $q(\gamma) = \tilde{q}(t) = q_0 = {\rm const}$, equation~(\ref{eq-kin-general-2}) becomes an Euler equation:
\beq
 f'(\gamma) +  {1\over{\beta_A}} {f\over\gamma}\, [c_1 + c_2  q_0 ] -
c_3 \beta_A \biggl(  q_0\gamma f' \biggr)' =0 \, , 
\label{eq-kin-general-alpha=2_v1}
\eeq
while the corresponding equation~(\ref{eq-kin-general-4}) simplifies to a linear homogeneous ordinary differential equation with constant coefficients:
\beq
 \dot{f} (t) +   f(t)\, [c_1 + c_2  q_0] - c_3 q_0 \, \ddot{f}  =0 \, . 
\label{eq-kin-general-alpha=2_v2}
\eeq
This can be readily solved as 
\beq
f(t) = C_1 e^{\lambda_1 t} + C_2 e^{\lambda_2 t} \, , 
\label{eq-soln-general-alpha=2}
\eeq
where $C_1$, $C_2$ are arbitrary constants and  $\lambda_1$ and $\lambda_2$ are the roots of the characteristic polynomial
\beq
P(\lambda) = \lambda^2 - {1\over{c_3 q_0}} \lambda - {{c_1+c_2 q_0}\over{c_3 q_0}} \, .
\label{eq-char_polynomial-alpha=2}
\eeq
For illustration, for the case $c_1 = c_2 = c_3 = 1$ considered in this paper, these roots are
$\lambda_1 =-1$, $\lambda_2 = 1+1/q_0$, and the corresponding solution is 
\beq
f = C_1 e^{\lambda_1 t} + C_2 e^{\lambda_2 t}  = 
C_1 \gamma^{\lambda_1/\beta_A} + C_2 \gamma^{\lambda_2/\beta_A} = 
C_1 \gamma^{-{1\over{\beta_A}}} + C_2 \gamma^{{1+q_0}\over{q_0 \beta_A}} \, .
\eeq
Imposing the condition that $f(\gamma)$ be a declining function of $\gamma$, we have to discard the second solution and thus recover the result $f(\gamma) \sim \gamma^{-1/\beta_A}$ we obtained previously, see~(\ref{eq-p-alpha=2_withFermi}). 
We note that the solution $\lambda_1 =-1$, $f(\gamma) \sim \gamma^{-1/\beta_A}$ remains valid even more generally, as long as $c_1 = 1$ and $c_2=c_3$.






\section{Summary and Conclusions}
\label{sec-conclusions}

In this paper we presented a physical picture and developed a theoretical model for nonthermal acceleration of relativistic particles in two-dimensional magnetic reconnection (with no guide field) in the limit of very large system sizes where reconnection takes place in the plasmoid-dominated regime. We focus here on the population of energetic particles in the active reconnection zone, before they are trapped inside plasmoids. The main premise of this work is that, just as one should approach describing the dynamics of reconnection in the plasmoid regime in a self-similar manner, viewing a large-scale reconnection layer as a self-similar hierarchical chain of plasmoids with a broad range of sizes (\S~\ref{subsec-picture-chain}), one should apply an analogous self-similar reasoning to the problem of particle acceleration. The motion of a highly energetic particle is not strongly sensitive to electromagnetic field structures on scales much smaller than the particle's Larmor radius. 
Thus, the motion --- and acceleration --- of each such particle should be analyzed by looking at it on the corresponding scale, i.e., by blurring the complex electromagnetic field structure of the reconnecting plasmoid chain to the scale commensurate with the given particle's characteristic Larmor radius. In particular, this means ignoring the interaction of this particle with small plasmoids, which may constitute the majority of the plasmoid population. A given energetic particle effectively sees a truncated plasmoid hierarchy and interacts only with plasmoids comparable or larger than its Larmor radius.  Since the Larmor radius of a relativistic particle is directly proportional to the particle's energy, higher-energy particles interact with a smaller number of large plasmoids, hence covering larger distances between such interactions while being continuously accelerated by the main reconnection electric field. Thus, the properties of nonthermal particle acceleration are fundamentally tied to the statistical characteristics of the plasmoid population, namely, to the plasmoid distribution function. 

In particular, particle acceleration in our picture (\S~\ref{subsec-picture-particle-acceleration}) is governed by the interplay between direct acceleration by the electric field $E_z$ in the ignorable ($z$) direction and particle deflection towards the $x$-direction (the direction along the reconnecting magnetic field) and magnetization by the reconnected magnetic field~$B_y$, followed by the particle's eventual trapping inside a large plasmoid.  We argue that in a self-similar reconnection layer these field components are approximately scale-invariant over a broad range of scales.  The accelerating $E_z$ electric field has two principal components, reflecting the nonlinear dynamical development of two instabilities acting in the layer. 
The first one is the (secondary) tearing, or plasmoid, instability, almost synonymous with the reconnection process itself; it provides the main reconnection electric field, $-E_{\rm rec} \hat{z}$, in the inter-plasmoid reconnection layers at every step in the hierarchy and leads to the growth of plasmoids of all sizes. 
The second instability is the plasmoid coalescence instability that drives rapid (up to the Alfv\'en speed) plasmoid motions along the layer in the $\pm x$ direction, and thus generates intense double-layer spikes of the associated motional electric field; in its linear stage this instability acts as a secondary, parasitic instability with respect to the tearing mode because it feeds upon the growing plasmoids. This motional electric field can accelerate or decelerate energetic particles as they collide and bounce off a large moving plasmoid by executing about one half of a cyclotron orbit in the plasmoid's $B_y$ magnetic field. The result of a single such encounter depends on the relative orientation between the particle's and the plasmoid's $x$-components of motion: the particle gains energy if the collision is head-on and loses energy if the collision is tail-on. The overall net effect after many such collisions can be described as Fermi acceleration: in general, it is a mixture of second-order diffusive acceleration (present if the motions of different plasmoids are uncorrelated), and rapid first-order acceleration if a given particle bounces many times between two approaching plasmoids before they finally collide and merge with each other; in either case, particles gain energy on average.

The reconnected magnetic field $B_y$ also plays an important role: it limits the acceleration of particles by deflecting them away from the direction of the accelerating electric field $E_z$ and eventually magnetizing them onto cyclotron orbits. While in a real reconnecting plasmoid chain this field can have a rather complex spatial structure (along the $x$-axis), here for simplicity we represent it as being roughly bimodal, at any scale in the self-similar hierarchy. Namely, we regard $|B_y|$ as being of order $B_1 =\epsilon B_0 \sim 0.1 B_0$ in the inter-plasmoid current layers and being of order $B_0$ in fully formed and circularized plasmoids. More precisely, $B_y$ varies (roughly linearly) from $-\epsilon B_0$ to $+\epsilon B_0$ in the $x$-direction along the midplane  of any given inter-plasmoid current layer, reversing sign as it passes through that layer's X-point; and it varies from $+B_0$ to $-B_0$ across a plasmoid, reversing sign as it passes through the O-point. At any given moment of time the reconnected field lines in the inter-plasmoid current layers are moving roughly Alfv\'enically towards an adjacent plasmoid, where they will eventually join the plasmoids' magnetic flux and compress.  Both types of the reconnected magnetic field exert a negative effect on particle acceleration by magnetizing energetic particles and thus preventing them from moving freely in the $z$ direction along the accelerating electric field. Thus, the reconnected field effectively removes energetic particles from the active acceleration zone, eventually placing them into quarantine inside large plasmoids --- large enough to confine them. 

In the mathematical formulation of the proposed theory (see \S~\ref{sec-model})  the main object of study is the energy distribution function $f(\gamma)$ of energetic particles present in the active acceleration zone, i.e., in interplasmoid current layers (which themselves can be plasmoid-dominated plasmoid chains), not yet trapped inside plasmoids. Each of the physical processes discussed above is represented by a separate term in the kinetic equation for $f(\gamma)$ [see~(\ref{eq-kinetic-1})]. Thus, the steady, regular acceleration by the main reconnecting electric field $E_{\rm rec}$ is described by the energy-advection term~$-\partial_\gamma (\dot{\gamma}_{acc} f)$; the 2nd-order Fermi acceleration due to particle reflections off moving plasmoids is described by the energy-diffusion term~$\partial_\gamma (D_\gamma \partial_\gamma f)$ (for simplicity we ignore 1st-order Fermi acceleration due to a particle bouncing back and forth between two converging plasmoids); 
the removal of particles from the active acceleration zone by their magnetization by the reconnected magnetic field $B_y$ is described by the escape term~$-f/\tau(\gamma)$. The latter is further subdivided into two contributions: due to the magnetization by the $\mathcal{O}(B_1=\epsilon B_0)$ reconnected field in the layers,  $-f/\tau_{\rm magn}(\gamma)$, and due to particle trapping inside large plasmoids, $-f/\tau_{\rm trap}(\gamma)$.  In principle, there is also a source term at small energies that describes the injection of particles into the reconnection region from the upstream region; however, assuming that the upstream plasma is magnetically dominated, so that its plasma-$\beta$ parameter is small, this term can be neglected in the supra-thermal high-energy range of interest to this study. 

By analyzing the resulting kinetic equation in a steady state, we find that different terms influence the key characteristics of the resulting nonthermal particle distribution --- e.g., its power-law index $p$ and its high-energy cutoff $\gamma_c$ --- in different ways. Thus, if one ignores the diffusive Fermi-type acceleration by moving plasmoids, then the balance between the direct, regular acceleration by $E_{\rm rec}$ and magnetization by the reconnected magnetic field~$B_1$ controls the power-law index of the particle spectrum: $p(\sigma_h) \simeq \beta_A^{-1} = (1+\sigma_h^{-1})^{1/2}$ [see~(\ref{eq-p-magn})].  Particle trapping by large plasmoids, on the other hand, mostly affects the high-energy part of the distribution, governing the high-energy cutoff. The functional shape of the cutoff and the value of the cutoff energy~$\gamma_c$ are controlled by the details of the plasmoid size (or flux) distribution function~$F(w)$ (see \S~\ref{subsec-trapping}). For example, if the plasmoid spectrum is a truncated power law, $F(w \leq w_{\rm max}) \sim w^{-\alpha}$ [with $\alpha\in(1,2)$], then the high-energy cutoff of the particle spectrum is exponential-like, $\sim \exp[-(\gamma/\gamma_c)^\nu]$, with $\nu = 2-\alpha$ and $\gamma_c$ determined by the truncation size of~$F(w)$ in combination with other parameters such as $\beta_A$ and~$\alpha$ (see \S~\ref{subsec-plasmoid_chain-simple} for details). The case $\alpha=2$ is, however, special: instead of an exponential cutoff, in this case one gets a steepening of the particle power law, with the net spectral index~$p$ increasing by a factor of order unity relative to what it would be just due to the magnetization by $B_1$ alone [see~(\ref{eq-p_tot-alpha=2})]. We also consider a more general and realistic case of a double power-law plasmoid distribution (see~\S~\ref{subsec-real-chain}).

The inclusion of the 2nd-order diffusive Fermi acceleration due to particle bouncing off randomly moving plasmoids raises the order of the differential kinetic equation and thus increases the mathematical complexity of the problem (see \S~\ref{subsec-general_kin_eq}).  We find however (\S~\ref{subsec-Fermi}) that for a single power-law plasmoid distribution with an index $1<\alpha<2$, the diffusive term is small for most the particles and is only important for particles near the maximum energy $\gamma_{\rm max}$ that can be confined only by the largest plasmoids, of size~$w_{\rm max}$. 
Interestingly, in the special case $\alpha=2$ the full problem, including the diffusive Fermi acceleration effects, can be solved exactly; the Fermi acceleration in this case effectively cancels the effect of particle trapping by large plasmoids, producing a remarkably simple solution $f \sim \gamma^{-p}$ with $p \simeq \beta_A^{-1}$ --- the same as would result without both of these plasmoid-related effects.

\smallskip
The theory developed in this paper represents one of the first steps towards building comprehensive statistical understanding of relativistic NTPA in 2D magnetic reconnection in the complex large-system, plasmoid-dominated regime, with simultaneous inclusion of several intertwined acceleration and escape channels. However, it is, admittedly, incomplete. 
First, even within the idealized 2D, purely anti-parallel (zero guide field)  scenario, the conceptual model outlined in this article ignored the population of particles already trapped inside plasmoids and, in particular, did not consider their possible subsequent acceleration due to the contraction and circularization of newly added plasmoid flux surfaces, or due to the slow, adiabatic compression of plasmoid cores, or due to plasmoid mergers. The model also did not account for the first-order Fermi acceleration due to particles bouncing multiple times between two large plasmoids that are steadily approaching each other and eventually merging; in this case, the changes in the particle energy due to subsequent encounters with the plasmoids are correlated and thus result in a fast, monotonic energy gain, in contrast with the random walk energy evolution in the case of uncorrelated plasmoid motions, which results in slower, diffusive 2nd-order Fermi acceleration considered in this article.  All these additional acceleration processes will need to be incorporated into the theory in future studies. 

Moreover, the presence of a finite magnetic field component in the $z$ direction --- either due to the preexisting guide magnetic field or the quadrupole Hall magnetic field that arises spontaneously in reconnection taking place in collisionless electron-ion plasmas --- can substantially modify the particles' trajectories and thus affect the dynamics and relative importance of the various acceleration and escape processes discussed in this paper.  An additional potentially important aspect of the Hall effect in electron-ion reconnection is the development, concurrently with the Hall quadruple magnetic field, of a bipolar in-plane ($xy$) electrostatic electric field; this electric field might not be strong and extended enough to significantly affect highly nonthermal particles in the high-energy tail of the distribution (which is the focus of this article) but it may affect the injection of particles into this tail from the thermal bulk.  
Furthermore, in the case of relativistic reconnection ($\sigma_h \gg 1$), a finite guide magnetic field $B_g$, in addition to its direct effect on particles trajectories, can also influence the acceleration process indirectly: its enthalpy, $B_g^2/4\pi$, adds to the enthalpy of the plasma, and thus increases the overall relativistic inertia of the fluid that needs to be moved by the reconnection outflows. 
This effectively reduces the relevant Alfv\'en velocity and slows down all plasma motions (including plasmoid motions), hence reducing the accelerating $E_z$ electric field, inhibiting particle acceleration.  In addition, the pressure of the guide field inside plasmoids reduces the compressibility of the plasmoid cores, thereby limiting the effectiveness of adiabatic particle acceleration in the plasmoids. 

Finally, the present model is purely 2D, which greatly restricts its ability to describe reconnection-driven particle acceleration in real 3D plasmas.  Three-dimensional effects can modify particle acceleration in a variety of ways. For example, 3D instabilities, such as the drift-kink, can destroy the neatly organized, nested flux surfaces that comprise 2D magnetic islands/plasmoids (or their 3D generalizations --- flux ropes). This allows the particles that would be permanently trapped inside plasmoids in 2D to escape along chaotic 3D field lines back into the upstream region. This giving the particles a new chance to be accelerated again, thus enhancing particle acceleration, especially in the presence of a modest guide field.  On the other hand, a strong guide field suppresses the drift-kink instability, and 3D effects in general, thus making a 3D reconnection layer more structurally similar to its 2D counterpart and, in particular, inhibiting particle's ability to escape from flux ropes and reenter the active acceleration zones. Thus, a guide magnetic field may play a nontrivial role in 3D reconnection-driven particle acceleration. 

All these processes and complex aspects of the problem will need to be investigated in greater detail further. 
We hope that the theoretical ideas expressed in this paper can provide some useful guidance in building a more comprehensive and realistic theory.  But the main tool for developing our physical understanding of particle acceleration in reconnection will undoubtedly be computational---namely, PIC simulations.  Numerical studies are rapidly becoming more powerful, opening the doors to rigorous and systematic ab initio exploration of reconnection in very large 2D and 3D systems where reconnection takes place in the plasmoid-mediated, turbulent regime. Increasing computing power also enables us to study systematically broader parameter spaces, e.g., allowing us to investigate the effects of guide field, system size (in principle, in all 3 directions separately), upstream plasma beta and magnetization, plasma composition (i.e., pair, electron-ion, or mixed-composition plasmas), and even more exotic radiative and QED regimes. In addition, because of the inherently stochastic nature of the large-system, plasmoid-dominated reconnection regime, one will need to conduct statistical ensembles of nearly identical simulations to probe the effects of stochasticity and characterize random variability of these processes. Finally, in addition to numerical and analytical studies, one may expect a strong growth of experimental contributions to our understanding of reconnection-driven particle acceleration in the coming years, including those from both traditional (magnetic confinement-based), and high-energy-density (laser-plasma and pulsed-power) experiments. 

The result of these efforts will be a thorough characterization of reconnection-driven NTPA that will enable us to formulate useful, practical prescriptions that can be reliably extrapolated to real astrophysical system sizes and thus applied to high-energy astrophysical objects.


\section*{Acknowledgements}

I am very grateful to many colleagues for numerous insightful discussions of reconnection-driven particle acceleration we had over the years, especially to Greg Werner, Mitch Begelman, Lorenzo Sironi, Maria Petropoulou, Jim Drake, and Nuno Loureiro. This work was supported by DOE grants DE-SC0008409, NASA grants NNX16AB28G and NNX17AK57G, and NSF grants AST-1411879, AST-1806084, and  AST-1903335.



\bibliographystyle{jpp}
\bibliography{recn-NTPA}

\end{document}